\documentclass[
reprint, prd, aps, twocolumn, nofootinbib, preprintnumbers, superscriptaddress, balancelastpage, longbibliography,
 amsmath,amssymb,
 aps,
]{revtex4-2}

\newcommand{\Fermi}{\textit{Fermi}}
\newcommand{\Msun}{M$_\odot$}

\newcommand{\beq}{\begin{equation}}
\newcommand{\eeq}{\end{equation}}

\newcommand{\mnras}{Monthly Notices of the Royal Astronomical Society}
\newcommand{\apjs}{The Astrophysical Journal Supplement Series}

\newcommand{\apjl}{The Astrophysical Journal Letters}
\newcommand{\aap}{Astronomy \& Astrophysics}
\newcommand{\aapr}{Astronomy and Astrophysics Reviews}
\newcommand{\apss}{Astrophysics and Space Science}

\newcommand\orcid[1]{\href{https://orcid.org/#1}{$\!$\includegraphics[scale=0.0045]{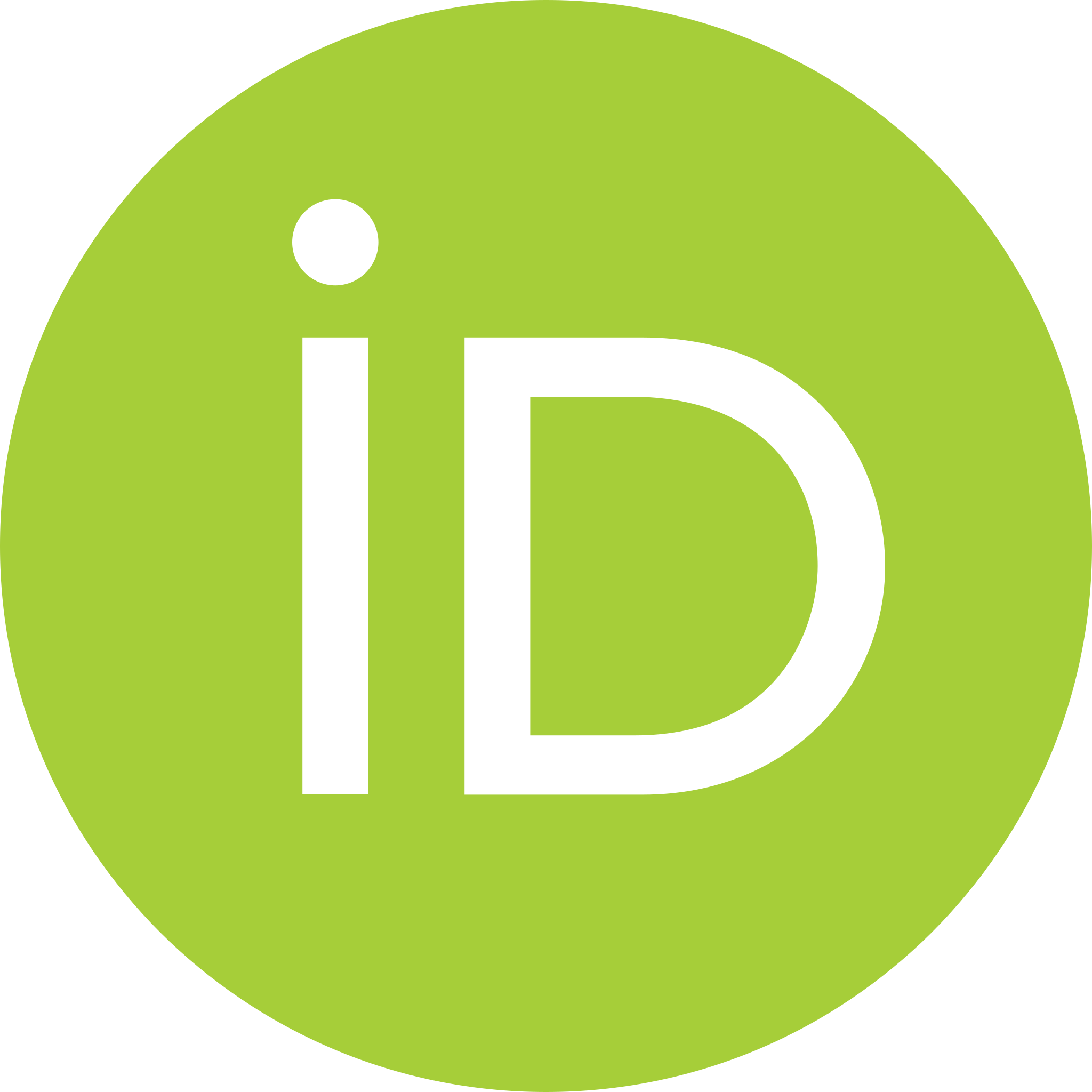} $\!\!$}}

\usepackage{graphicx}
\usepackage{dcolumn}
\usepackage{bm}
\usepackage{hyperref}
\usepackage{color}
\usepackage{adjustbox}
\usepackage{longtable}
\usepackage{multirow}
\usepackage{float}
\usepackage{tabularx}
\usepackage{array} 
 
\usepackage{threeparttable}
\usepackage[dvipsnames]{xcolor}
\definecolor{pomegranate_purple}{HTML}{b90078}

\hypersetup{
    colorlinks   = true,
    citecolor    = pomegranate_purple,
    urlcolor     = pomegranate_purple,
    linkcolor    = pomegranate_purple
}

\begin{document}

%\preprint{APS/123-QED}

\title{Are X-Ray Detected Active Galactic Nuclei in Dwarf Galaxies Gamma-Ray Bright?}

\author{Milena Crnogor\v{c}evi\'{c} \orcid{0000-0002-7604-1779}}\email{milena.crnogorcevic@fysik.su.se}
\affiliation{The Oskar Klein Centre, Department of Physics, Stockholm University, Stockholm 106 91, Sweden}

\author{Tim Linden \orcid{0000-0001-9888-0971}}\email{linden@fysik.su.se}
\affiliation{The Oskar Klein Centre, Department of Physics, Stockholm University, Stockholm 106 91, Sweden}
\affiliation{Erlangen Centre for Astroparticle Physics (ECAP), Friedrich-Alexander-Universität \\ Erlangen-Nürnberg, Nikolaus-Fiebiger-Str. 2, 91058 Erlangen, Germany}

\author{Annika H.~G.~Peter \orcid{0000-0002-8040-6785}}\email{peter.33@osu.edu}
\affiliation{Center for Cosmology and Astroparticle Physics, The Ohio State University, 191 West Woodruff Avenue, Columbus, OH 43210, USA}
\affiliation{Department of Physics, The Ohio State University, 191 West Woodruff Avenue, Columbus, OH, 43210, USA}
\affiliation{Department of Astronomy, The Ohio State University, 140 West 18th Avenue, Columbus, OH 43210, USA}

\date{\today}% It is always \today, today,

\begin{abstract}
The $\gamma$-ray emission from active galactic nuclei (AGN), including both beamed blazars and misaligned-AGN, dominates the extragalactic $\gamma$-ray point-source population count and flux. While multi-wavelength studies have detected an increasing number of AGN within dwarf galaxies in the local Universe, $\gamma$-ray emission has so far only been associated with systems hosting supermassive black holes (SMBHs). Dwarf-galaxy AGN are of particular interest because their central black holes fall in the intermediate-mass black hole (IMBH) regime, offering insight into the early evolution of SMBHs. Using 15~years of \textit{Fermi}-LAT data, we present the first search for $\gamma$-ray emission from dwarf-galaxy AGN. In the sample of 74 X-ray-selected dwarf-galaxy AGN, we find no sources that exceed the \textit{Fermi}-LAT detection threshold. However, a joint-likelihood analysis reveals a modest, trials-corrected population-level excess ($\sim2\sigma$) above blank-field expectations at very soft photon indices $\Gamma \gtrsim 3.8$ above 500~MeV. This hint is most pronounced when source contributions are weighed by $M^\alpha_{{\rm IMBH},i}/d_i^2$, with $\alpha\simeq1$--$1.5$, suggesting---but not confirming---that $\gamma$-ray emission could scale with the central black hole mass or a property correlated with it (e.g., accretion rate), but with a markedly softer spectrum than in SMBH-hosted AGN.

\end{abstract}

\maketitle

\section{Introduction}
\label{sec:intro}
\vspace{-0.2cm}
Intermediate-mass black holes (IMBHs) form the missing link in the black hole mass spectrum. With masses between $10^4$ and $10^6$~\Msun, IMBHs occupy a critical---yet poorly understood---regime between stellar-mass black holes ($\sim$10~\Msun) and supermassive black holes (SMBHs, $\gtrsim$ $10^6$~\Msun). IMBHs are central to understanding black hole formation, growth, and their co-evolution with host galaxies \citep{Miller:2003sc, Greene2020, 2017IJMPD..2630021M}. Recent surveys have revealed that IMBHs are not rare: more than half of dwarf galaxies with $M_* > 10^7$~\Msun\ may host a central IMBH \cite{Burke:2024wcf, 2023MNRAS.523.5610B, 2025ApJ...985..223M, 2025ApJ...982...10P}. 

\begin{figure}[!t]
    \centering
    \includegraphics[width=0.9\linewidth]{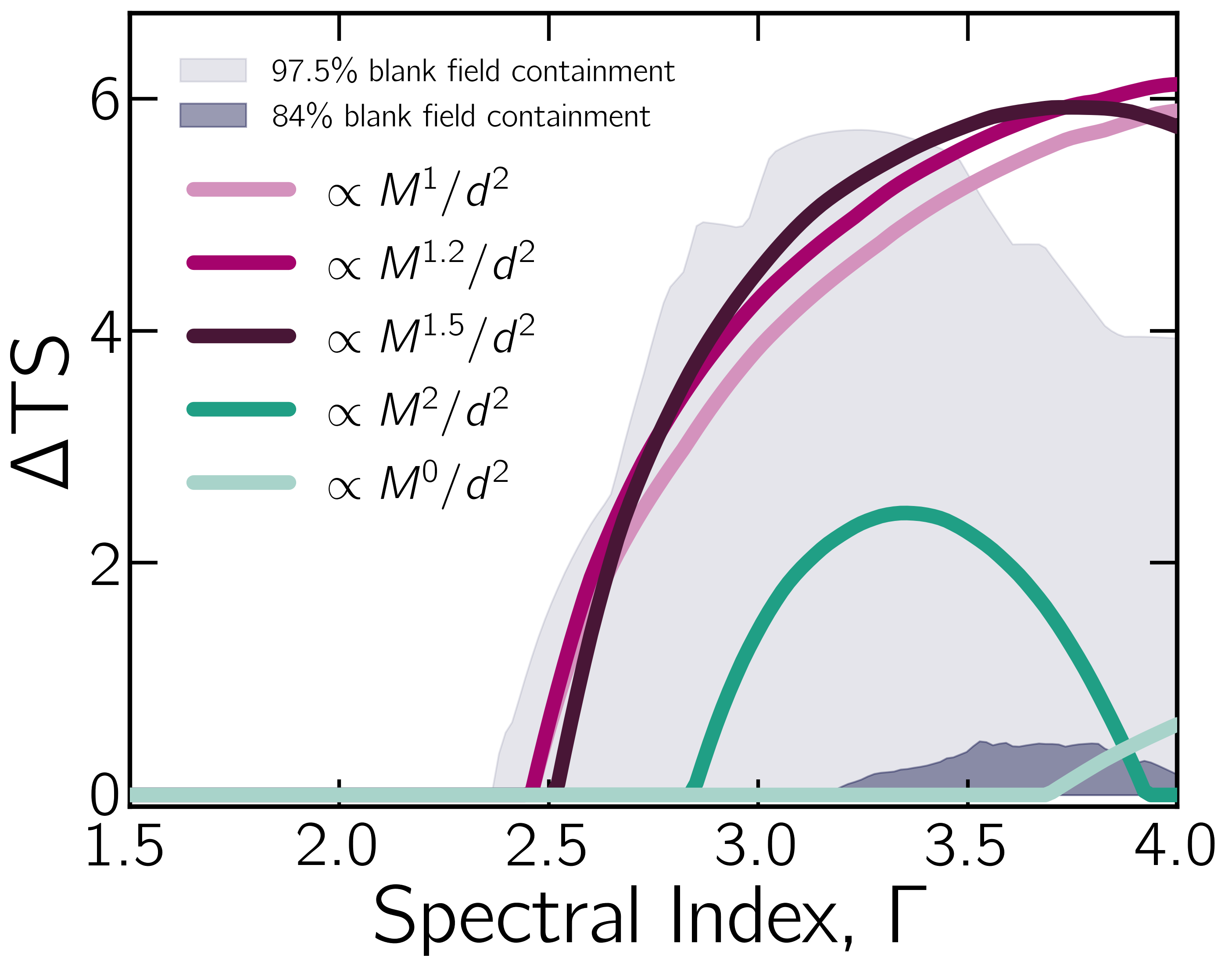}
    \caption{Joint-likelihood TS profiles ($\Delta$TS) as a function of photon index $\Gamma$ for weights $w_i \propto M_i^{\alpha}/d_i^{2}$ for x-ray selected dwarf AGN. Colored lines show $\alpha=0$ (distance only), $1$, $1.2$, $1.5$, and $2$. Shaded bands indicate the 84\% (dark) and 97.5\% (light) containment of blank-field profiles. Weightings with near-linear mass dependence ($\alpha \sim 1$–$1.5$) yield the highest $\Delta$TS values, peaking at soft spectra $\Gamma \sim 3.8$–$4.0$. By contrast, distance-only weighting ($\alpha=0$) and strong mass-up weighting ($\alpha=2$) remain consistent with the background.}
    \label{fig:alpha_scan}
\end{figure}

Central black holes (both IMBHs and SMBHs) play a pivotal role in the evolution of their hosts from the earliest cosmic epochs to the present day \citep{1998SilkRees, Croton:2005hbr, 2010A&ARv..18..279V, Kormendy:2013dxa, Heckman:2014kza, Reines:2016kej, 2019MNRAS.489..802R}. Moreover, as central black holes accrete mass and grow, they produce bright multi-wavelength emission that is potentially detectable at cosmological distances. In the case of SMBHs, observations of high-redshift active galactic nuclei (AGN), which are powered by the accretion onto the central black hole, reveal that SMBHs with masses reaching $10^9$~\Msun\ already existed less than a billion years after the Big Bang \citep{SDSS:2005dqi, Willott:2007rm, Mortlock:2011va, Venemans:2013npa, DES:2015ija}. Recent JWST results have extended these discoveries to even earlier epochs, when the Universe was only about 500 million years old \citep{Larson:2023, Goulding:2023gqa, Bogdan:2023ilu}.

The presence of billion-solar-mass black holes less than a billion years after the Big Bang imposes stringent constraints on black hole ``seed'' formation mechanisms. Theoretical models propose two primary pathways: ``light seeds'' \mbox{($\sim$10--100~\Msun)} formed from the collapse of Population III stars, and ``heavy seeds'' \mbox{($\sim10^4$--$10^6$~\Msun)} formed through the direct collapse of pristine gas clouds \cite{Madau:2001sc,  Bromm:2002hb, 2006MNRAS.371.1813L, 2010A&ARv..18..279V, Latif:2013pyq, Inayoshi:2019fun, Greene2020}. Crucially, cosmological simulations show that dwarf galaxy black hole demographics are uniquely sensitive to the seed formation mechanism \cite{Chadayammuri:2022bjj, 2025MNRAS.538..518B}. Unlike massive galaxies where subsequent growth can erase seeding signatures, the lower-mass environments and limited growth in dwarf galaxies preserve imprints of the original seed population \cite{Burke:2024wcf, 2025MNRAS.538..518B}. As a result, searches for IMBHs in dwarf galaxies provide some of the most direct observational tests of black hole seed formation and early growth \citep{Reines:2013pia, Lemons:2015ara, Baldassare:2016cox, Reines:2022ste, Ansh:2022tmn, Sanchez:2024gbl}.

% DESI, can remove
Recent multi-wavelength surveys have dramatically expanded our census of AGN in dwarf galaxies. The MaNGA integral field spectroscopy survey identified 664 AGN in dwarf galaxies, revealing an AGN fraction of $\sim$20\%---significantly higher than previous estimates of $\sim$0.5--1\% from single-fiber studies \cite{2024MNRAS.528.5252M}. Simultaneously, the Dark Energy Spectroscopic Instrument (DESI) early data release identified 2,444 dwarf-galaxy AGN (hereafter referred to as dwarf AGN) candidates \cite{2025ApJ...982...10P}. These advances demonstrate that AGN activity is far more common in dwarf galaxies than previously recognized. Cosmological simulations support these observational findings, predicting that many IMBHs may be off-center or remain undetected by current surveys due to low luminosities and stellar contamination \cite{Bellovary:2018gbb, 2022ApJ...936...82S,  2023ApJ...957...16S}.

% why x-rays
X-ray observations provide particularly robust AGN identification in dwarf galaxies, where optical emission line diagnostics fail to identify 85\% of x-ray selected sources due to contamination from star formation processes \cite{2020MNRAS.492.2268B}. For $\gamma$-ray follow-up studies, the x-ray sample of dwarf AGN comprises the nearest sources and enables a robust blank field calibration for statistical analysis.
X-rays may also offer a physical connection to $\gamma$-ray activity: in massive galaxies, multi-wavelength studies of blazars have revealed a moderate correlation between x-ray and $\gamma$-ray luminosities (e.g., \cite{Fossati:1998zn, 2011MNRAS.414.2674G}), reflecting their common origin in relativistic jet populations, with about one-third of x-ray bright BL Lacs detected by \textit{Fermi}-LAT \cite{Fermi-LAT:2022byn}. In unbeamed AGN systems, however, the correlation is weaker: $\gamma$-ray activity tends to track radio jet power more closely than coronal x-rays, and most Seyfert-like systems are $\gamma$-ray quiet. Thus, our choice of x-ray selection is motivated primarily by its robustness for dwarf AGN identification, while the presence or absence of correlated $\gamma$-ray emission remains an open question.

The eROSITA All-Sky Survey has revolutionized population-level x-ray astronomy, increasing the number of known x-ray sources by more than 60\% and providing the statistical foundation for population studies \cite{eROSITA:2024oyj}. Early pilot studies of dwarf galaxies with eROSITA identified six AGN candidates with x-ray luminosities $L_{0.5-8~\text{keV}} \sim 10^{39}$--$10^{40}~$erg s$^{-1}$, suggesting that when extrapolated to the full survey, eROSITA could detect $\sim$1,350 AGN candidates in dwarf galaxies \cite{2021ApJ...922L..40L}. 

% why gamma rays 
While x-ray observations can confirm the presence of accreting IMBHs, they provide limited information about the physical processes that drive high-energy emission. $\gamma$-ray observations offer complementary insights that can distinguish between different emission mechanisms and probe the fundamental physics of accretion and jet formation in lower-mass systems. Several theoretical models predict detectable $\gamma$-ray emission from IMBHs through distinct physical processes: (1) inverse-Compton scattering of ambient photons by relativistic electrons in coronae or jets \cite{Inoue:2019fil, Murase:2023ccp}, (2) synchrotron self-Compton emission from structured jets \cite{2008ApJ...686..181F, 2011MNRAS.413.1845T, Potter:2015tya, deMenezes:2020rah}, (3) cosmic-ray interactions producing neutral pions that decay into $\gamma$-rays \cite{Yang:2018dsi}, and (4) potentially, dark matter annihilation in density spikes around IMBHs \cite{1998SilkRees, Bertone:2005hw, Ferrer:2017xwm}. Observationally, $\gamma$-ray emission has been established in low-luminosity AGN (LLAGN) hosted by massive galaxies through several of these channels \cite{deMenezes:2020rah, Murase:2023ccp}. Although LLAGN and IMBH-hosting AGN are distinct populations, their overlapping physics suggests that similar emission processes could also operate in IMBH systems. 

Several critical observational and theoretical questions emerge from this context.
Key unknowns include whether the $\gamma$-ray emission mechanisms observed in massive galaxy AGN scale down to IMBHs, and how the emission efficiency depends on the black hole mass, accretion rate, and host galaxy environment.
$\gamma$-ray observations of IMBH systems in dwarf galaxies can test the universality of high-energy processes across different mass scales and potentially distinguish between the aforementioned competing emission models.

%this paper
In this paper, we present the first systematic search for $\gamma$-ray emission from AGN in dwarf galaxies. We utilize 15 years of \textit{Fermi}-LAT data to measure $\gamma$-ray luminosities and spectra for individual sources from the eROSITA All-Sky Survey and conduct a comprehensive joint likelihood analysis of the complete dwarf AGN sample \cite{Sacchi:2024yic}. Our analysis employs scaling relationships to test theoretical predictions for the correlation between $\gamma$-ray luminosity and black hole masses. We compare our results to extensive blank field studies to assess the statistical significance of any detected signals and characterize the spectral properties of potential $\gamma$-ray emission from the dwarf AGN population. Using a joint-likelihood analysis, we find a modest, trials-corrected excess at very soft photon indices ($\Gamma \gtrsim 3.8$) above 500 MeV. This hint, shown in Fig.~\ref{fig:alpha_scan}, is most pronounced when source contributions are weighted by $M^\alpha_{\rm IMBH}/d^2$ with $\alpha\simeq1$–$1.5$.

%sections
The paper is organized as follows. In Section~\ref{sec:data-selection}, we describe the x-ray selected dwarf AGN sample and estimate IMBH masses using established scaling relations. Section~\ref{sec:lat} outlines our \textit{Fermi}-LAT data analysis methodology, including individual source analyses, joint likelihood techniques, and blank field calibration procedures. In Section~\ref{sec:results}, we present the results of our $\gamma$-ray analysis, including individual source upper limits and the statistical significance of any population-level signals. Finally, we interpret our findings in the context of theoretical emission models and their implications for IMBH physics in Section~\ref{sec:discussion}, and summarize our conclusions and suggest directions for future research. Appendix~\ref{app:excluded} describes our analysis of sources excluded by \citet{Sacchi:2024yic} as a control sample, while Appendix~\ref{app:100MeV} presents extended analyses down to 100 MeV to test the robustness of our spectral measurements.

%%%%%%%%%%%%%%%%%%%%%%%%%%%%%%%%%%%%%%%%%%%%%%%%%%%%
%%%%%%%%%%%%%%%%%%%%%%%%%%%%%%%%%%%%%%%%%%%%%%%%%%%%
%%%%%%%%%%%%%%% Data Selection %%%%%%%%%%%%%%%%%%%%%
%%%%%%%%%%%%%%%%%%%%%%%%%%%%%%%%%%%%%%%%%%%%%%%%%%%%
%%%%%%%%%%%%%%%%%%%%%%%%%%%%%%%%%%%%%%%%%%%%%%%%%%%%

\section{Data Selection and Characterization}
\label{sec:data-selection}

We conduct the analysis of the sample of x-ray-bright AGN in dwarf galaxies compiled by Sacchi \emph{et al.}~\citep{Sacchi:2024yic}, which used data from the first eROSITA all-sky survey (eRASS1). To ensure the robustness of this sample, the authors of Ref.~\citep{Sacchi:2024yic} (1) employ Monte Carlo simulations to rule out the possibility of chance alignments with background AGN, (2) apply scaling relations to exclude associations with stellar processes such as x-ray binary formation, and (3) use luminosity thresholds to identify and exclude contamination from ultra-luminous x-ray sources (ULXs), which typically also originate from stellar remnants. Sacchi \emph{et al.} argue that sources which significantly exceed these thresholds are highly likely to be AGN and are thus retained in the sample. After eliminating the most prominent contaminants, the authors finalize a sample that consists of 74 dwarf galaxy-IMBH pairs. We present the properties of these systems, along with the predicted IMBH mass and $\gamma$-ray flux, in Table~\ref{tab:prop}. Finally, for our $\gamma$-ray analysis, we apply additional quality cuts---detailed below---that reduce our final sample to 67 sources.

\noindent \textbf{Galactic Plane Sources.} --- We exclude six targets---PGC747791, HIZOAJ1343$-$65, PGC096512, ESO221$-$009, ESO271$-$018, and PGC538542---from the primary sample because they lie at low Galactic latitude ($|b| < 15^\circ$), where structured diffuse emission and source crowding dominate the systematics. Within each target's region of interest (RoI), several bright \emph{Fermi}-LAT sources lie within $\sim$1$^\circ$, making reliable background modeling challenging. 

\noindent \textbf{Confusion with Bright $\gamma$-ray Sources.} --- We exclude LAMOSTJ134630.99+070428.6 as an independent LAT target because it lies only 0.166$^{\circ}$ from the bright catalog source 4FGL J1345.8+0706, a flat-spectrum radio quasar (FSRQ) which fully accounts for the local photon excess \cite{Fermi-LAT:2022byn}. When considered individually, the apparent signal arises only if the 4FGL source is effectively held fixed, allowing the test source to absorb its flux. LAMOSTJ134630.99+070428.6 is amongst the most distant dwarfs in our sample and only moderately massive, so its exclusion is unlikely to have a significant impact on the joint-likelihood analysis. To avoid double-counting photons and artificially inflating the population signal, we therefore remove LAMOSTJ134630.99+070428.6 from the dwarf-AGN target list.

These exclusions leave us with a final sample of 67 dwarf AGN for our $\gamma$-ray analysis, all located at high Galactic latitude ($|b| > 15^\circ$) and free from contamination by nearby bright $\gamma$-ray sources. These sources are all listed in Table~\ref{tab:prop}.

\subsubsection*{Estimating Black Hole Masses in Dwarf AGN Sample}
To interpret the $\gamma$-ray properties of our dwarf AGN sample, we first estimate the masses of their central black holes. Because direct dynamical mass measurements are not available for these low-mass systems, we estimate the black hole mass ($M_{\mathrm{IMBH}}$) for each source using empirical scaling relations. Specifically, we adopt the relation from Reines \& Volonteri \cite{2015ApJ...813...82R}, which links black hole mass $M_{\mathrm{IMBH}}$ to host galaxy stellar mass $M_*$ through 
\begin{equation}
\label{eq:imbh_mass}
\log \left( \frac{M_{\mathrm{IMBH}}}{M_\odot} \right) = 7.45 + 1.05 \left[ \log \left( \frac{M_*}{10^{10} M_\odot} \right) \right].
\end{equation}

Stellar masses for the host galaxies are taken from Ref.~\cite{Sacchi:2024yic}, who derives them via spectral energy distribution (SED) fitting using stellar population synthesis models and a standard initial mass function (e.g., Salpeter or Chabrier). While this approach is standard, it carries uncertainties due to assumptions about the star formation history, dust attenuation, and metallicity. Applying Eq.~\ref{eq:imbh_mass} introduces further uncertainty, as the Reines \& Volonteri relation has an intrinsic scatter of $\sim$0.55 dex. We do not explicitly propagate this scatter into our $\gamma$-ray analysis; instead, we adopt the best-estimate values as a consistent, first-order proxy for $M_{\mathrm{IMBH}}$ across the sample. The resulting black hole masses are listed in Table~\ref{tab:prop}. 
 
\begin{figure*}[t]
\centering
\includegraphics[width=0.95\textwidth]{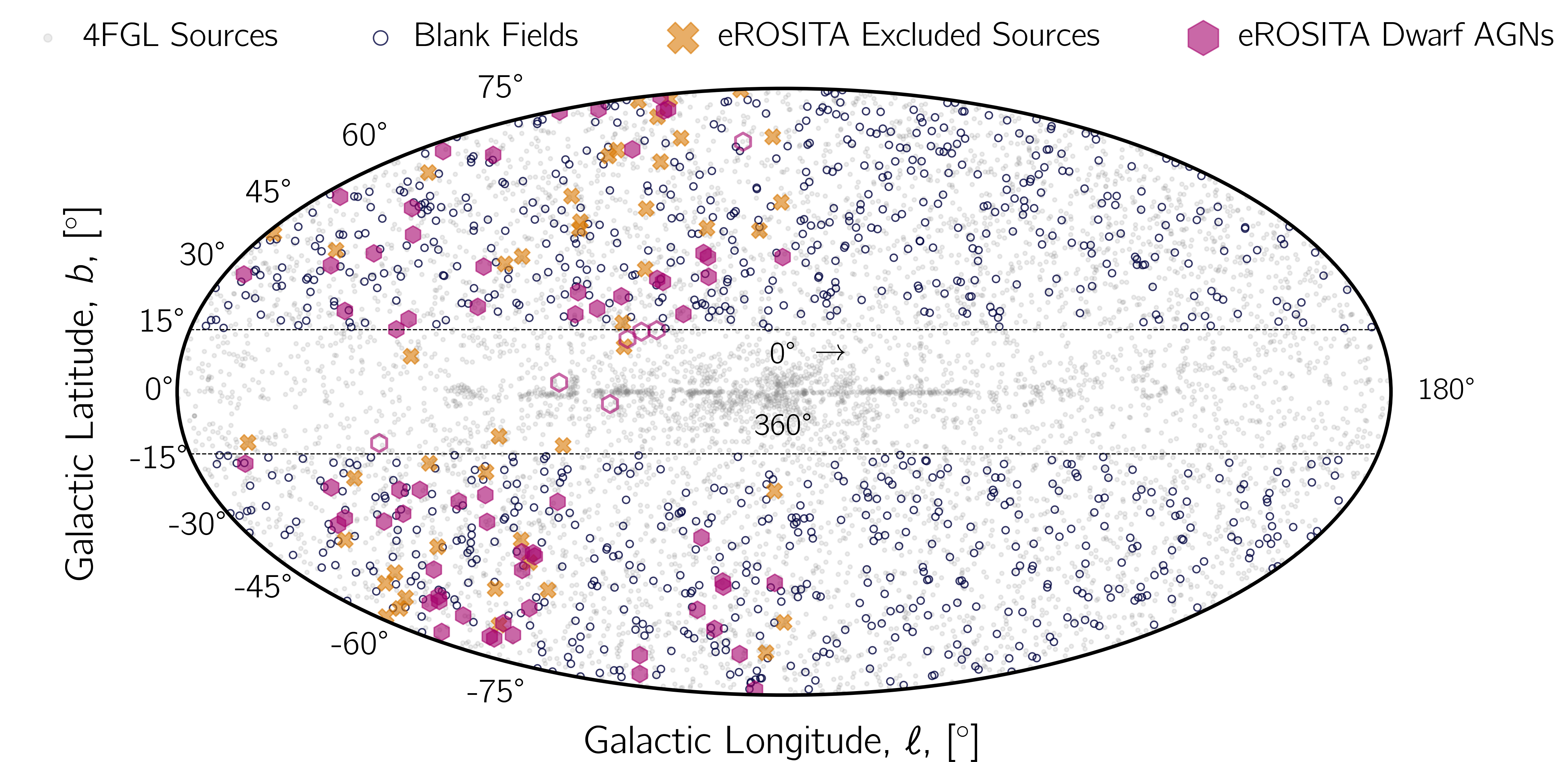}
\caption{Distribution of the 1000 blank fields (navy circles), the 74 eROSITA-selected dwarf AGN (magenta hexagons), and all sources in the 4FGL-DR4 catalog (light gray dots). Sources excluded from the AGN sample by \citet{Sacchi:2024yic} --- including ULXs, background AGN, and XRBs --- are shown as orange crosses (Appendix~\ref{app:excluded}). Empty magenta hexagons indicate the 7 dwarf AGN excluded from our $\gamma$-ray analysis: LAMOSTJ134630.99+070428.6 (confusion with bright 4FGL source) and 6 Galactic plane sources (PGC747791, HIZOAJ1343-65, PGC096512, ESO221-009, ESO271-018, PGC538542). Horizontal dashed lines mark $|b|=15^\circ$ in Galactic latitude. Our final $\gamma$-ray sample consists of 67 sources (filled hexagons).}
\label{fig:blanks}
\end{figure*}

\section{\Fermi-LAT data analysis}
\label{sec:lat}

The Large Area Telescope (LAT), onboard the \textit{Fermi} $\gamma$-ray observatory, detects $\gamma$-rays with energies ranging from $\sim$100 MeV to $\sim$500 GeV. The LAT has a field of view of 2.4 sr, covering approximately 20\% of the sky at a time, and offering source localization with angular resolution down to a few arcminutes. A detailed overview of the LAT instrument and its performance can be found in Ref. \cite{2009ApJ...697.1071A}. 

For this analysis, we use 15 years of \texttt{Pass 8} reprocessed data, covering the period from August 4, 2008, to August 4, 2023, and an energy range from 500~MeV to 500~GeV, obtained from the \textit{Fermi} Science Support Center (FSSC) website\footnote{\url{https://fermi.gsfc.nasa.gov/ssc/data/}, accessed on September 1, 2025.}. We adopt a minimum energy threshold of 500~MeV to mitigate the diffuse background systematics that dominate observations at lower energies. For completeness, we also extend our analysis down to 100~MeV, and present these results in the Appendix~\ref{app:100MeV}. We use the \texttt{fermipy} software package (v.~1.2.0) to analyze data, which relies on \texttt{FermiTools} (v.~2.2.0)\footnote{\url{https://fermi.gsfc.nasa.gov/ssc/data/analysis/software/}, accessed on September 1, 2025.} for likelihood-based modeling and fitting \cite{fermipyopensourcepythonpackage}. We employ standard analysis techniques similar to those used in previous LAT studies of dwarf spheroidal galaxies, but adapted to our sample of dwarf AGN (see, e.g.,~\cite{Fermi-LAT:2015att, Fermi-LAT:2016uux, Hoof:2018hyn, Linden:2019soa, McDaniel:2023bju, Crnogorcevic:2023ijs}). 

We select photons from the \texttt{P8R3\_SOURCE} event class. To reduce contamination from the Earth’s limb, we impose a zenith angle cut at 100$^{\circ}$ and apply the standard good time interval (GTI) selection. For each source, we define the RoI of $10^\circ \times 10^\circ$, centered on the AGN coordinates reported in Table~\ref{tab:prop}. To account for photon leakage from outside the RoI due to the LAT point spread function, we construct our source model over a larger $15^\circ \times 15^\circ$ area. In cases where multiple dwarf AGN fall within a single RoI, the central target is analyzed as the primary source, while the additional (non-central) dwarfs are included in the model as point sources with power-law spectra to avoid double-counting their potential contribution. The data are binned into eight logarithmic energy bins per decade, with a spatial bin size of $0.1^\circ$.

\begin{table*}
\centering
\caption{X-ray identified dwarf AGN compiled by Sacchi \emph{et al.}~\citep{Sacchi:2024yic}. 
Stellar masses ($M_*$) are derived from SED fitting \citep{Sacchi:2024yic}, and IMBH masses ($M_{\rm IMBBH}$) are estimated using the scaling relation of Reines \& Volonteri~\citep{2015ApJ...813...82R} (Eq.~\ref{eq:imbh_mass}). 
The X-ray luminosities ($L_X$) are taken directly from \citet{Sacchi:2024yic}, while the estimated $\gamma$-ray fluxes ($F_\gamma$) are obtained using Eq.~\ref{eq:x-ray-gamma-ray} from \cite{2013Ap&SS.347..349L}. 
\textit{Italicized rows indicate sources excluded from our analysis due to their location at low Galactic latitude or confusion with bright nearby $\gamma$-ray sources}, while \textbf{bold rows highlight the sources with the highest TS values in our individual source likelihood scans (Fig.~\ref{fig:spaghetti_tmax}).}}

\label{tab:prop}
\resizebox{\textwidth}{!}{%
\begin{tabular}{lcccccccc}
\hline\hline
\textbf{ID} & \textbf{R.A.} & \textbf{Dec} & \textbf{D [Mpc]} & \textbf{$\log(M*/M_\odot)$} & \textbf{$\log(M_{\rm IMBH}/M_\odot)$} & \textbf{$L_X$ [$10^{40}$ erg/s]} & \textbf{$F_\gamma$ [$10^{-10}$ ph cm$^{-2}$ s$^{-1}$]} \\
\hline
PGC135069 & 52.29 & -22.02 & 24.50 & 9.41 & 5.78 & 6.55 & 2.93 \\
WINGSJ043139.17-612330.1 & 67.91 & -61.39 & 41.03 & 8.22 & 4.53 & 11.37 & 2.40 \\
PGC200277 & 179.38 & 32.34 & 48.36 & 9.02 & 5.37 & 3.10 & 1.21 \\
6dFJ1551458-091534 & 237.94 & -9.26 & 75.19 & 8.87 & 5.21 & 24.03 & 1.97 \\
ESO411-012 & 11.71 & -31.54 & 19.71 & 7.66 & 3.94 & 0.22 & 0.85 \\
\textbf{PGC2822735} & \textbf{25.21} & \textbf{-32.74} & \textbf{166.27} & \textbf{9.45} & \textbf{5.82} & \textbf{64.99} & \textbf{1.54} \\
6dFJ0142532-432355 & 25.72 & -43.40 & 19.22 & 7.78 & 4.07 & 1.72 & 2.05 \\
NGC1034 & 39.56 & -15.81 & 19.28 & 9.15 & 5.51 & 0.27 & 0.94 \\
6dFJ0245228-240617 & 41.35 & -24.10 & 101.61 & 9.24 & 5.60 & 72.89 & 2.44 \\
ESO479-025 & 40.53 & -24.13 & 16.90 & 9.31 & 5.68 & 0.19 & 0.91 \\
PGC3213080 & 42.59 & -29.57 & 76.40 & 8.76 & 5.10 & 2.57 & 0.76 \\
PGC012154 & 49.09 & -25.85 & 19.66 & 9.37 & 5.74 & 0.25 & 0.90 \\
PGC693568 & 47.03 & -32.33 & 65.84 & 8.75 & 5.09 & 5.14 & 1.15 \\
PGC540963 & 50.55 & -44.09 & 122.51 & 9.28 & 5.64 & 3.66 & 0.59 \\
PGC134076 & 53.72 & -24.79 & 21.76 & 8.45 & 4.77 & 0.19 & 0.73 \\
6dFJ0342278-260243 & 55.62 & -26.05 & 21.40 & 8.23 & 4.54 & 0.18 & 0.73 \\
PGC3315038 & 63.10 & -54.42 & 173.09 & 9.47 & 5.84 & 4.92 & 0.50 \\
\textbf{6dFJ0421194-313805} & \textbf{65.33} & \textbf{-31.63} & \textbf{32.95} & \textbf{8.67} & \textbf{5.00} & \textbf{8.19} & \textbf{2.51} \\

PGC126927 & 66.46 & -61.15 & 68.36 & 8.88 & 5.22 & 0.75 & 0.50 \\
ESO118-034 & 70.07 & -58.74 & 13.65 & 9.27 & 5.63 & 0.21 & 1.13 \\
PGC928553 & 73.20 & -14.13 & 39.15 & 9.37 & 5.74 & 0.95 & 0.88 \\
PGC016675 & 76.13 & -16.58 & 43.61 & 9.37 & 5.74 & 0.59 & 0.66 \\
PGC016986 & 78.84 & -26.47 & 50.47 & 8.71 & 5.05 & 1.38 & 0.83 \\
UGC03282 & 79.41 & 6.80 & 80.21 & 9.42 & 5.79 & 3.47 & 0.83 \\
6dFJ0532495-322925 & 83.21 & -32.49 & 148.13 & 9.47 & 5.84 & 10.68 & 0.79 \\
PGC917425 & 84.06 & -14.98 & 26.12 & 8.82 & 5.16 & 0.62 & 1.03 \\
2MASXJ05403594-5436417 & 85.15 & -54.61 & 32.35 & 9.43 & 5.80 & 0.38 & 0.70 \\
PGC148318 & 90.96 & -33.19 & 36.89 & 9.17 & 5.53 & 0.54 & 0.73 \\
PGC075555 & 93.06 & -38.77 & 21.07 & 8.43 & 4.75 & 0.20 & 0.77 \\
\textbf{PGC483594} & \textbf{92.44} & \textbf{-48.80} & \textbf{122.17} & \textbf{9.27} & \textbf{5.63} & \textbf{10.04} & \textbf{0.91} \\
PGC019486 & 100.79 & -76.56 & 50.73 & 9.13 & 5.49 & 0.43 & 0.51\\ 
\textbf{PGC088580} & \textbf{97.75} & \textbf{-56.65} & \textbf{142.62} & \textbf{9.41} & \textbf{5.78} & \textbf{25.28} & \textbf{1.18} \\
\textit{PGC747791} & \textit{102.77} & \textit{-27.94} & \textit{31.13} & \textit{9.10} & \textit{5.46} & \textit{0.93} & \textit{1.06} \\
PGC2054090 & 120.44 & 34.76 & 65.89 & 9.31 & 5.68 & 2.13 & 0.80 \\
PGC023484 & 125.62 & -1.10 & 62.38 & 9.16 & 5.52 & 34.09 & 2.68 \\
PGC3093997 & 130.95 & -17.68 & 22.81 & 9.34 & 5.71 & 0.43 & 0.99 \\
PGC024469 & 130.70 & 14.27 & 29.99 & 9.45 & 5.82 & 0.44 & 0.80 \\
PGC153352 & 134.75 & -18.38 & 26.86 & 9.25 & 5.61 & 0.60 & 1.00 \\
NGC2777 & 137.67 & 7.21 & 25.70 & 9.33 & 5.70 & 0.74 & 1.13 \\
IC0549 & 145.18 & 3.96 & 22.70 & 8.93 & 5.28 & 0.45 & 1.02 \\
PGC2077238 & 145.83 & 36.24 & 87.52 & 9.46 & 5.83 & 4.91 & 0.89 \\
AGC193818 & 147.96 & 13.91 & 98.06 & 9.39 & 5.76 & 31.27 & 1.76 \\
NGC3125 & 151.64 & -29.94 & 14.98 & 9.37 & 5.74 & 0.49 & 1.49 \\
6dFJ1024202-201458 & 156.08 & -20.25 & 69.73 & 8.96 & 5.31 & 15.63 & 1.76 \\
IC2604 & 162.35 & 32.77 & 25.10 & 8.97 & 5.32 & 0.44 & 0.92 \\
UGC05989 & 163.13 & 19.79 & 16.07 & 8.78 & 5.12 & 0.51 & 1.43 \\
\textit{PGC096512} & \textit{173.70} & \textit{-59.23} & \textit{18.51} & \textit{8.48} & \textit{4.80} & \textit{0.13} & \textit{0.72} \\
NGC4016 & 179.62 & 27.53 & 50.62 & 9.46 & 5.83 & 1.60 & 0.88 \\
ESO321-018 & 183.98 & -38.09 & 41.79 & 9.14 & 5.50 & 0.93 & 0.83 \\
ESO267-032 & 183.69 & -43.56 & 26.01 & 9.22 & 5.58 & 0.22 & 0.67 \\
2MASXJ12260880+2826008 & 186.54 & 28.43 & 64.95 & 9.12 & 5.48 & 2.16 & 0.81 \\
PGC039730 & 184.97 & 1.77 & 29.37 & 8.72 & 5.06 & 0.39 & 0.77 \\
\hline\hline
\end{tabular}%
}
\end{table*}

\begin{table*}
\centering
\addtocounter{table}{-1} % keep the same table number
\caption{Continued.}
\resizebox{\textwidth}{!}{%
\begin{tabular}{lcccccccc}
\hline\hline
\textbf{ID} & \textbf{R.A.} & \textbf{Dec} & \textbf{D [Mpc]} & \textbf{$\log(M*/M_\odot)$} & \textbf{$\log(M_{\rm BH}/M_\odot)$} & \textbf{$L_X$ [$10^{40}$ erg/s]} & \textbf{$F_\gamma$ [$10^{-10}$ ph cm$^{-2}$ s$^{-1}$]} \\
\hline
PGC039904 & 185.30 & 17.64 & 16.98 & 8.69 & 5.02 & 0.18 & 0.88 \\
UGC07366 & 184.87 & 17.23 & 17.52 & 8.92 & 5.27 & 0.15 & 0.80 \\
PGC560327 & 191.99 & -42.65 & 26.50 & 9.13 & 5.49 & 0.24 & 0.68 \\
ESO324-008 & 200.22 & -39.28 & 27.12 & 8.99 & 5.34 & 0.66 & 1.03 \\
\textit{HIZOAJ1343-65} & \textit{205.85} & \textit{-65.26} & \textit{43.61} & \textit{8.94} & \textit{5.29} & \textit{0.88} & \textit{0.78} \\
\textit{LAMOSTJ134630.99+070428.6} & \textit{206.63}& \textit{7.07} & \textit{110.43} & \textit{9.17} & \textit{5.53} & \textit{33.71} & \textit{1.65} \\
\textit{ESO221-009} & \textit{207.69} & \textit{-48.95} & \textit{41.13} & \textit{8.92} & \textit{5.27} & \textit{1.20} & \textit{0.93} \\

\textbf{NGC5398} & \textbf{210.34} & \textbf{-33.06} & \textbf{11.39} & \textbf{9.21} & \textbf{5.57} & \textbf{0.23} & \textbf{1.37} \\
\textit{ESO271-018} & \textit{212.52} & \textit{-46.22} & \textit{35.29} & \textit{8.15} & \textit{4.46} & \textit{0.52} & \textit{0.75} \\
PGC679199 & 212.61 & -33.23 & 50.80 & 9.40 & 5.77 & 0.97 & 0.71 \\
\textit{PGC538542} & \textit{218.22} & \textit{-44.32} & \textit{15.51} & \textit{8.65} & \textit{4.98} & \textit{4.72} & \textit{3.75} \\
ESO580-005 & 219.57 & -22.32 & 34.14 & 9.06 & 5.41 & 0.59 & 0.81 \\
PGC812038 & 221.43 & -22.58 & 47.16 & 9.46 & 5.83 & 13.42 & 2.29 \\
PGC053603 & 225.12 & -26.45 & 74.12 & 9.46 & 5.83 & 7.49 & 1.22 \\
6dFJ1458243-373012 & 224.60 & -37.50 & 104.80 & 9.33 & 5.70 & 64.95 & 2.27 \\
PGC329372 & 309.74 & -63.77 & 21.35 & 8.44 & 4.76 & 0.29 & 0.89 \\
2MASXJ21390659-4353001 & 324.78 & -43.88 & 80.99 & 9.47 & 5.84 & 22.01 & 1.79 \\
PGC3321235 & 332.55 & -56.07 & 55.31 & 8.61 & 4.94 & 5.99 & 1.43 \\
PGC3320749 & 330.19 & -56.69 & 63.14 & 8.81 & 5.15 & 2.61 & 0.90 \\
NGC7657 & 351.70 & -57.81 & 40.41 & 9.37 & 5.74 & 0.74 & 0.77 \\
ESO240-012 & 354.71 & -51.86 & 23.69 & 9.41 & 5.78 & 0.41 & 0.94 \\
ESO241-006 & 359.06 & -43.43 & 18.94 & 8.92 & 5.27 & 0.29 & 0.98 \\
\hline\hline
\end{tabular}%
}
\end{table*}

For the Galactic diffuse emission, we use the standard interstellar emission model (\texttt{gll\_iem\_v07}), while the isotropic diffuse background is modeled using \texttt{iso\_P8R3\_SOURCE\_V3}. We include both point and extended sources within the RoI using the 4FGL-DR4 catalog (\texttt{gll\_psc\_v35}), allowing for dispersion corrections (with \texttt{edisp=True}) \cite{Fermi-LAT:2022byn}. The parameters for the Galactic diffuse emission and isotropic diffuse background are left free to vary. We also free the normalizations of all sources with a test statistic (TS) greater than 25 that lie within $5^\circ$ of the RoI center, as well as those within $7^\circ$ and TS$>$500. We model all dwarf AGN sources as point-like, justified by the AGN's small angular sizes and the LAT's angular resolution limits \cite{2009ApJ...697.1071A}. We also note that, because eROSITA selection is insensitive to jet orientation, any $\gamma$-ray signal we uncover is not contingent on beaming; our analysis formalism treats each target as an (unresolved) point source regardless of orientation.

\subsubsection*{Individual Source Likelihood Analysis} 
\label{subsec:individual}
After constructing the RoI model, we apply a binned likelihood fit in each energy bin. We test for $\gamma$-ray emission by comparing models with and without a source at the AGN position. The significance of any signal is measured using $TS = -2~{\rm log} \left(\mathcal{L}_0/\mathcal{L}_1\right)$, where $\mathcal{L}_0$ and $\mathcal{L}_1$ are the likelihoods for the null and source-included models, respectively. For each AGN, we compute the SED by fixing the per–energy–bin spectral index to a common value, $\Gamma_{\rm bin}=2$, and fitting only the normalization in each bin. This produces bin–by–bin flux points that are independent of any assumed global spectrum. The differential flux in each bin is parametrized as  
\begin{equation}
    \left(\frac{d\Phi_\gamma}{dE}\right)_{\rm bin} = c \left(\frac{E}{E_0}\right)^{-\Gamma_{\rm bin}},
\end{equation}
\noindent
where $E_0$ denotes the reference energy in all bins (here defined as $E_0=1~\rm{GeV}$). To estimate the SED, we construct a likelihood profile \(\mathcal{L}\left(\frac{d\Phi_\gamma}{dE}, E\right)\) and then sum it over individual energy bins, 
\begin{equation}
\label{eq:likelihood}
\log\mathcal{L} = \sum_{E_i} \log\mathcal{L}\left(\frac{d\Phi_\gamma}{dE}\left(E_i\right), E_i\right),
\end{equation}
using the \texttt{fermipy}'s \texttt{gta.sed()} routine. We distinguish between the fixed per-bin slope $\Gamma_{\rm bin}=2$, adopted solely for SED extraction, and the global spectral index $\Gamma$, which is the free parameter scanned in the subsequent analysis. To map the likelihood across the power-law parameter space, we evaluate $\mathcal{L}$ on a grid of normalization and spectral index values, $(c, \Gamma)$. Specifically, we scan $c$ over $\log_{10}$-spaced values from $10^{-18}$ to $10^{-12}$ cm$^{-2}$ s$^{-1}$ (using 100 steps), and $\Gamma$ linearly from 1.5 to 4 (also in 100 steps). For each grid point, we compute the predicted power-law flux in the given energy bin, interpolate the likelihood from the SED fit, and sum over all bins to obtain the total likelihood as a function of $(c, \Gamma)$. Finally, we reference the resulting 2D likelihood surface to its global maximum by computing the difference $\Delta\log\mathcal{L} = \log\mathcal{L}(c, \Gamma) - \log\mathcal{L}_\mathrm{max}$. We use the test statistic, $TS = -2\Delta\log\mathcal{L}$ to determine the best-fit spectral parameters and derive confidence intervals.
% Remove or comment out these conflicting lines from your preamble:
% \usepackage{array}
% \newcolumntype{P}[1]{>{\raggedright\arraybackslash}p{#1}}
% \newcolumntype{M}[1]{>{\centering\arraybackslash}m{#1}}

% Replace the problematic table with this REVTeX-compatible version:

% Remove or comment out these conflicting lines from your preamble:
% \usepackage{array}
% \newcolumntype{P}[1]{>{\raggedright\arraybackslash}p{#1}}
% \newcolumntype{M}[1]{>{\centering\arraybackslash}m{#1}}

% Replace the problematic table with this REVTeX-compatible version:

% Remove or comment out these conflicting lines from your preamble:
% \usepackage{array}
% \newcolumntype{P}[1]{>{\raggedright\arraybackslash}p{#1}}
% \newcolumntype{M}[1]{>{\centering\arraybackslash}m{#1}}

% Replace the problematic table with this REVTeX-compatible version:

% Remove or comment out these conflicting lines from your preamble:
% \usepackage{array}
% \newcolumntype{P}[1]{>{\raggedright\arraybackslash}p{#1}}
% \newcolumntype{M}[1]{>{\centering\arraybackslash}m{#1}}

% Replace the problematic table with this REVTeX-compatible version:

% Remove or comment out these conflicting lines from your preamble:
% \usepackage{array}
% \newcolumntype{P}[1]{>{\raggedright\arraybackslash}p{#1}}
% \newcolumntype{M}[1]{>{\centering\arraybackslash}m{#1}}

% Replace the problematic table with this REVTeX-compatible version:

\begin{table*}[t]
\caption{Representative expectations for the mass--only scaling $L_\gamma \propto M^\alpha$ at fixed Eddington ratio. ``Hard'' denotes $\gamma$-ray spectral index in $\Gamma\sim 2$--$2.5$ range; and ``soft'' denotes $\Gamma\gtrsim 3$.}
\label{tab:alpha_summary}
\begin{ruledtabular}
\begin{tabular}{clllr}
$\alpha$ & Physical scenario & Spectral expectation in $\gamma$ rays & Typical sources & Key refs. \\
\hline
1.0 & Accretion/Eddington--tracked & Broad possibilities; often \emph{soft} if & Radiatively efficient AGN & Refs. \\
    & power (disk, corona, BZ/MAD jets) & coronal/RIAF Comptonization dominates & (Seyferts, some NLS1) & \cite{Merloni:2003aq, Tchekhovskoy2011, Ghisellini:2014pwa, 2009ApJ...699..976A, Kimura:2020thg} \\
    &  & can be harder if jet IC dominates & jetted AGN & \\[1ex] \hline
$\sim$1.0 & Hadronic $pp$ in RIAFs/hot flows & Typically \emph{hard} in GeV band & LLAGN/LINERs, Sgr~A*-like & Refs. \\
    &  & ($\Gamma\sim 2$--$2.5$) unless strong & nuclei, nearby radio galaxies & \cite{2013MNRAS.432.1576N, Kelner:2006tc} \\
    &  & cooling/absorption steepens spectrum & in RIAF states & \\[1ex] \hline
$\sim$1.2 & Scale-invariant jets, optically & Synchrotron to $\gtrsim$MeV is & HBL/FR~I systems with & Refs. \\
    & thin synchrotron branch & steep/soft; GeV emission more & high synchrotron peaks; & \cite{Heinz:2003xt, Falcke:2003ia, 1992ApJ...397L...5M, 1994ApJ...421..153S} \\
    &  & likely IC from same leptons & jet-dominated LLAGN & \\[1ex] \hline
$\sim$1.5 & Scale-invariant jets, optically & Flat/inverted at low energies; & Compact flat-spectrum cores & Refs. \\
    & thick (flat) synchrotron branch & not expected to extend efficiently & (FR~I/II, radio-loud AGN) & \cite{Heinz:2003xt, Heinz:2004uu} \\
    &  & to GeV via synchrotron (typically IC) & & \\[1ex] \hline
$\gtrsim$1 & Super-Eddington/slim-disk & Often \emph{soft} due to strong & ULXs, some NLS1s, TDEs & Refs. \\
    & phenomenology (beaming/geometry; & Comptonization/winds; details & in super-Eddington phases & \cite{1988ApJ...332..646A, Kaaret2017ARAandA, Kitaki2017PASJ} \\
    & mass-dependent $\lambda_{\rm Edd}$) & model-dependent & & \\[1ex]\hline
0 & Control case (no mass scaling) & No specific spectral prediction & Baseline (null scaling) & -- \\
\end{tabular}
\end{ruledtabular}

\vspace{0.5em}
\raggedright
\footnotesize
\textit{\textbf{Acronyms:}} \textbf{BZ}: Blandford--Znajek; \textbf{FR~I/II}: Fanaroff--Riley type I/II radio galaxy; \textbf{HBL}: High-frequency-peaked BL Lac object; \textbf{IC}: Inverse Compton; \textbf{LINER}: Low-Ionization Nuclear Emission-line Region; \textbf{MAD}: Magnetically Arrested Disk; \textbf{NLS1}: Narrow-line Seyfert 1 galaxy; \textbf{RIAF}: Radiatively Inefficient Accretion Flow; \textbf{TDE}: Tidal Disruption Event.
\end{table*}

\subsubsection*{Joint Likelihood Analysis of the Sources}
\label{subsec:joint}
To maximize sensitivity to faint $\gamma$-ray emission associated with IMBHs in our sample of dwarf AGN, we implement a joint likelihood analysis that combines data from all 67 AGN in our sample\footnote{Joint likelihood is distinct from ``stacking,'' although the terms are often used interchangeably in the field. Stacking simply sums the counts and exposures of individual sources, neglecting variations in background and source brightness, whereas joint likelihood rigorously accounts for these differences. For faint populations, joint likelihood is the statistically correct approach.}.
This method, commonly used in \textit{Fermi}-LAT studies of faint source population (see, e.g., \cite{Fermi-LAT:2016uux, McDaniel:2023bju}) allows us to search for a cumulative signal that may otherwise be too weak to be detected in individual AGN. The joint likelihood is defined as the product of the individual likelihoods for each source (i.e., the sum of their log-likelihoods), expressed as:
\begin{equation} 
\label{eq:joint_likelihood} 
\log\mathcal{L}_{\text{total}}= \sum_i \log\mathcal{L}_i(\alpha_s, \alpha_i | D_i), 
\end{equation}
\noindent where \(\mathcal{L}_i\) is the likelihood for the \(i\)-th source, \(\alpha_s\) represents parameters shared across all sources (e.g., spectral index \(\Gamma\)), \(\alpha_i\) are source-specific nuisance parameters (e.g., flux normalization \(c_i\)), and \(D_i\) refers to the observational data for the \(i\)-th source. We maximize the joint likelihood function (Eq.~\ref{eq:joint_likelihood}) to determine the best-fit parameters for the combined signal. This allows us to construct profile likelihoods for the flux normalization $c$ and the spectral index $\Gamma$. The TS for the combined analysis is defined as 
\begin{equation} 
\text{TS} = -2\log \left(\mathcal{L}_{\text{total}, 0}/\mathcal{L}_{\text{total}, 1} \right), 
\label{eq:ts_combined}
\end{equation}
where $\mathcal{L}_{\text{total}, 0}$ is the likelihood under the null hypothesis (no signal) and $\mathcal{L}_{\text{total}, 1}$ corresponds to that of the alternative hypothesis. A key advantage of this joint-likelihood approach (compared to traditional counts stacking) is that it allows for a self-consistent test of theoretically predicted models for the relative flux from each source. In practice, each source's normalization is tied to a single shared amplitude through physically motivated weights, $c_i=Aw_i$, with $w_i$ encoding, for example, (i) purely geometric dimming ($w_i\propto 1/d_i^2$), or (ii) mass-distance scalings ($w_i\propto M_{\mathrm{IMBH},i}^\alpha/d_i^2$) incorporating the black hole mass raised to a power $\alpha$. 

We note that the parameter $\alpha$ is best interpreted as an \textit{effective} scaling: in most theoretical models, the primary driver of luminosity is the accretion rate, often expressed via the Eddington ratio, $\lambda_{\mathrm{Edd}}$. The black hole mass thus sets the characteristic physical scales, such that a simple power-law dependence $M^\alpha$ folds together (i) the intrinsic dependence of the emission mechanism on $M_{\mathrm{IMBH}}$ at fixed $\lambda_{\mathrm{Edd}}$, (ii) any systematic dependence of $\lambda_{\mathrm{Edd}}$ on $M_\mathrm{IMBH}$ across the sample (due to astrophysics or observational selection), and (iii) possible mass or accretion-rate dependence on the $\gamma$-ray radiative efficiency. Nevertheless, theoretical scenarios motivate characteristic ranges of $\alpha$, all summarized in Table~\ref{tab:alpha_summary}. For accretion- or Eddington–tracked power, one expects $\alpha \sim 1$~\cite{Merloni:2003aq, Tchekhovskoy2011, Ghisellini:2014pwa, 2009ApJ...699..976A, Kimura:2020thg}. Hadronic $pp$ emission in radiatively inefficient accretion flows (RIAFs) similarly gives an effective $\alpha \sim 1$, though typically with harder spectra ($\Gamma \sim 2{-}2.5$; \cite{2013MNRAS.432.1576N, Kelner:2006tc}). Scale-invariant jets predict mass scalings for synchrotron, $F_\nu\!\propto\!M^{17/12-\alpha_\nu/3}$ at fixed $\dot m$, resulting in  $\alpha\!\approx\!1.2$ on the optically thin branch and $\alpha\!\approx\!1.5$ on the optically thick (flat) branch~\citep{Heinz:2003xt, Falcke:2003ia, 1992ApJ...397L...5M, 1994ApJ...421..153S, Heinz:2004uu}. Finally, super-Eddington flows can exceed $L_{\mathrm{Edd}}$ by factors of a few, but still scale roughly with $M$; any apparent $\alpha > 1$ would most likely reflect a mass dependence of $\lambda_{\mathrm{Edd}}$ or of the radiative efficiency rather than a universal scaling~\citep{1988ApJ...332..646A, Kaaret2017ARAandA, Kitaki2017PASJ}. In this framework, $\alpha$ provides a phenomenological way to confront different physical hypotheses with the data---analogous to, for example, the $J$-factor scalings in dark matter searches in dwarf spheroidal galaxies.

Finally, we measure the significance of any observed signal by comparing the measured TS values to null distributions derived from two control samples: 1) blank fields, which account for instrumental and diffuse modeling uncertainties, and 2) sources excluded by Sacchi et al. \cite{Sacchi:2024yic}, such as x-ray binaries, ULXs, and background AGN, which helps validate the robustness of our AGN sample (Appendix~\ref{app:excluded}). 

\subsubsection*{Blank Fields}

Assessing the significance of any $\gamma$-ray excess requires a well-characterized background. While standard statistical assumptions---such as expecting the TS to follow a Poisson-like distribution---can break down in low-count regimes or for one-sided hypotheses, in practice the dominant deviation from Poisson expectations arise from imperfectly known backgrounds, {\it e.g.,} diffuse mismodeling, unresolved sources, and instrumental systematics \cite{Fermi-LAT:2013sme}.
%Standard statistical assumptions --- such as expecting the TS to follow a Poisson-like distribution --- often break down in low-count regimes or when testing one-sided hypothesis, resulting in TS distributions that differ substantially from expectations \cite{Fermi-LAT:2013sme}. 
To account for this, we construct an empirical background using 1000 high-latitude regions ($|b|>15^\circ$), excluding regions within the 95\% containment of 4FGL-DR4 sources and within 0.1$^\circ$ from known emitters listed in BZCAT, CRATES, and WIBRaLS blazar catalogs \cite{Carlson:2014nra, 2015Ap&SS.357...75M, Massaro:2008ye, Healey:2007by, 2014ApJS..215...14D}. The resulting distribution of blank-sky fields is shown in Fig.~\ref{fig:blanks}. We then apply the same analysis pipeline outlined in previous subsections. For each weighting mode (no weights, $1/d^2$, $M^\alpha_{\mathrm{IMBH}}/d^2$), we generate 400 blank-sky realizations by randomly selecting $N=67$ distinct blanks and assigning AGN-like physics weights via sampling of the AGN $(M_{\rm IMBH},d)$ pool with replacement. Finally, this yields a TS distribution that captures both statistical fluctuations and the modeling uncertainties, as well as instrumental systematics intrinsic to the \textit{Fermi}-LAT data.

\section{Results}
\label{sec:results}

\subsection{Individual Source Analysis}
\label{subsec:individual_results}
\begin{figure}[t]
\centering
\includegraphics[width=0.45\textwidth]{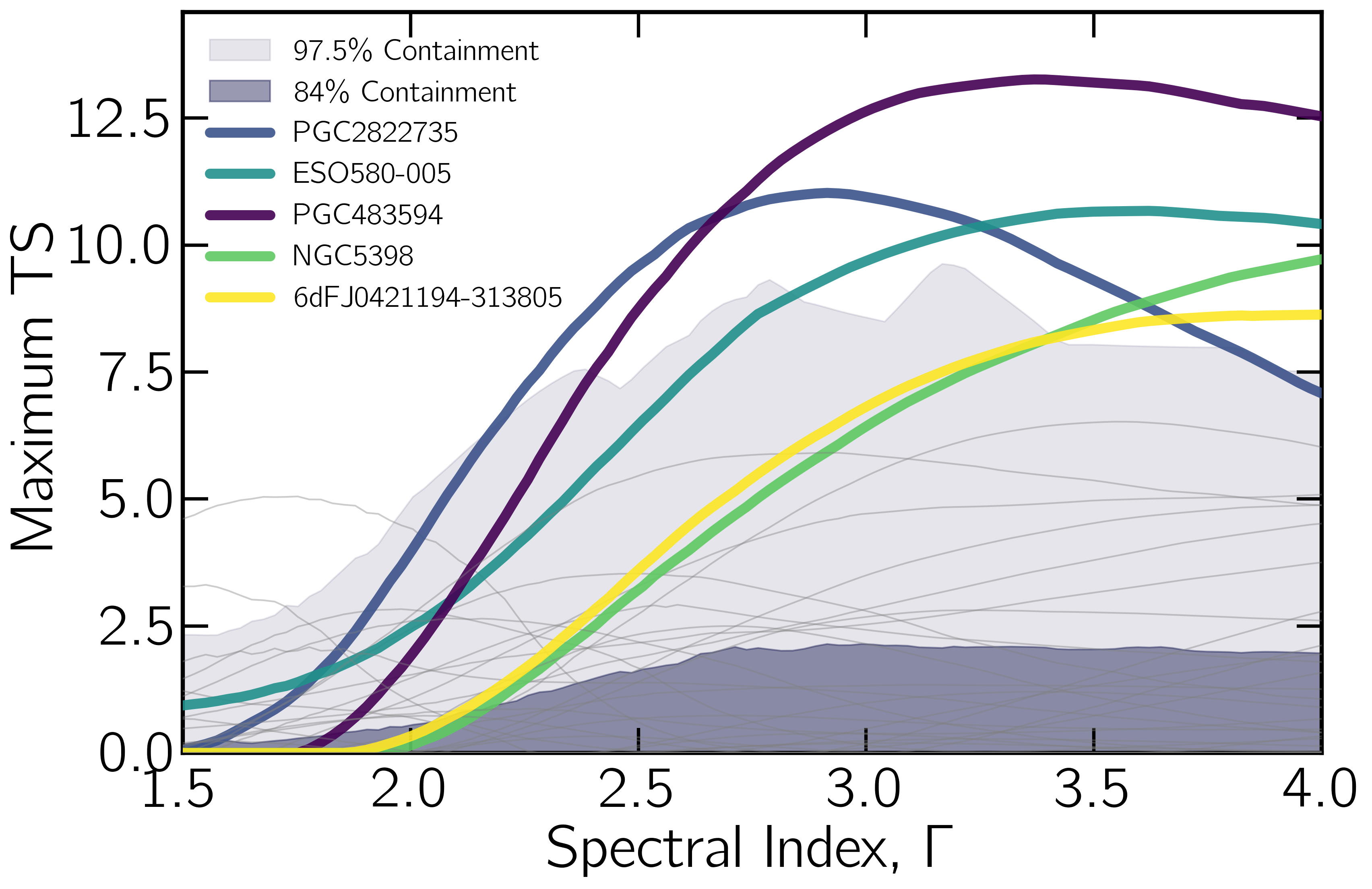}
\caption{Maximum TS as a function of spectral index $\Gamma$ over all flux normalizations for individual dwarf AGN candidates in comparison with blank field observations. Five sources with highest significances are shown as colored lines. The light gray and dark gray band show the 97.5\% and 84\% containment regions for the individual blank fields, respectively.}
\label{fig:spaghetti_tmax}
\end{figure}

We find no individual sources that exceed the typical \textit{Fermi}-LAT detection threshold of TS$>25$. The highest individual TS values range from $\sim$5--15, consistent with statistical fluctuations and blank field expectations given that  we are testing 67 sources. Figure~\ref{fig:spaghetti_tmax} shows the maximum TS values over all normalization values as a function of $\Gamma$ for the individual dwarf AGN, compared with the containment bands of the individual blank fields. 
% add note on whether these sources are the closest / the most massive.

\begin{figure}[t]
\centering
\includegraphics[width=0.45\textwidth]{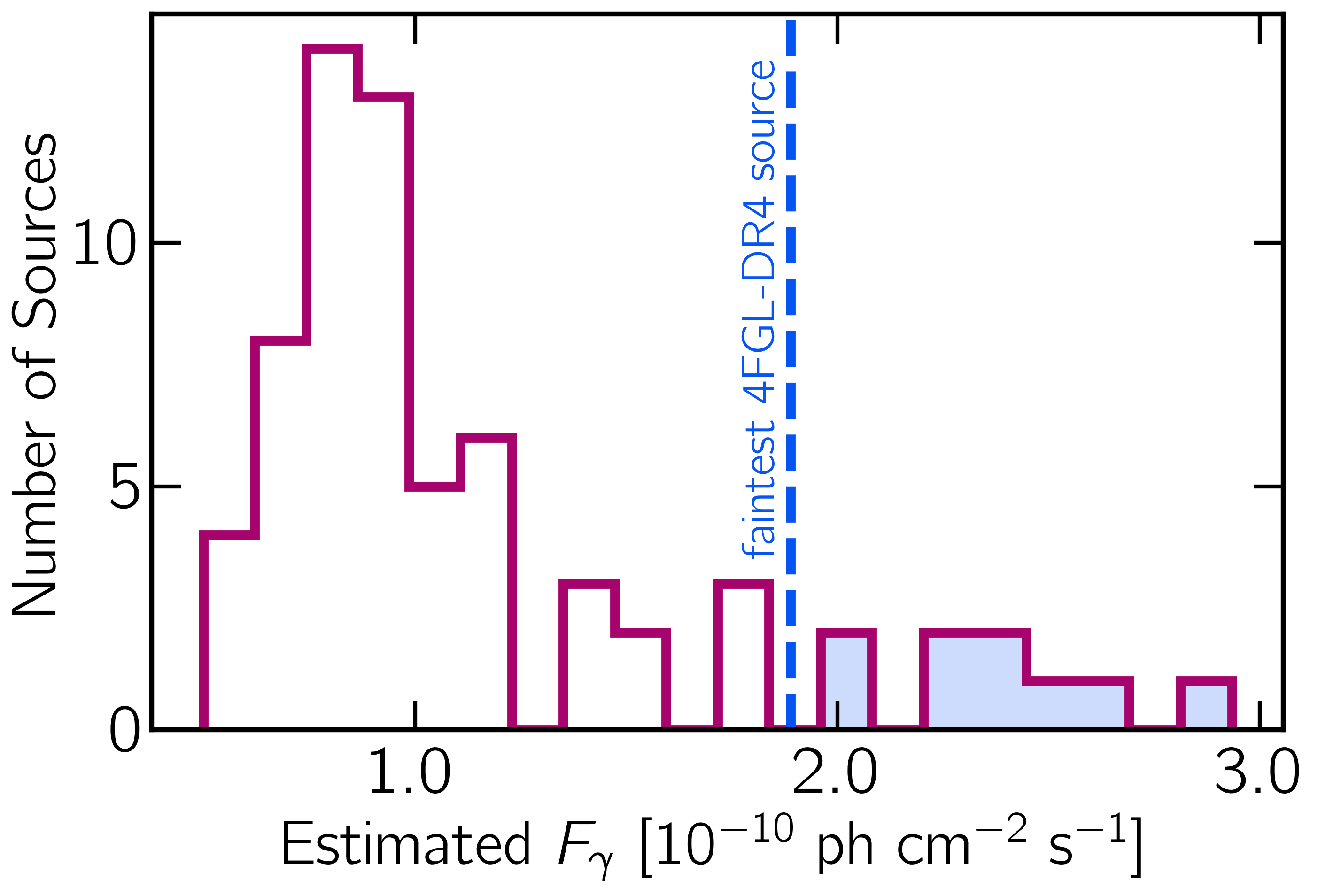}
\caption{Estimated $\gamma$-ray fluxes for the 67 eROSITA-selected dwarf AGNs based on their x-ray luminosities. The vertical dashed line marks the faintest 4FGL-DR4 source; shadowed bins indicate sources exceeding this threshold and could be individually detectable with \textit{Fermi}-LAT.}
\label{fig:gamma_flux_hist}
\end{figure}

To place these non-detections in context, we apply x-ray/$\gamma$-ray scaling relations to our sample from the known populations of BL Lac objects, where synchrotron self-Compton (SSC) processes dominate the high-energy emission \cite{Fossati:1998zn, 2011MNRAS.414.2674G}. While previous studies have shown only a moderate correlation between x-ray and $\gamma$-ray luminosities for blazars, a non-negligible fraction---about one-third---of x-ray bright BL Lacs are detected by \textit{Fermi}-LAT. Assuming a similar correlation holds for dwarf AGN, we estimate the expected $\gamma$-ray fluxes using the observed x-ray fluxes and the empirical x-ray–$\gamma$-ray relation derived for BL Lacs,
\begin{equation}
\label{eq:x-ray-gamma-ray}
\log F_{\gamma} = 0.42 \log F_{X} - 8.17,
\end{equation}
% \begin{equation}
% \label{eq:x-ray-gamma-ray}
% \log L_{\gamma} = (0.59 \pm 0.05) \log L_{X} + (17.55 \pm 2.31),
% \end{equation}
where both luminosities are in erg s$^{-1}$ and represent time-averaged (intermediate-state) values \cite{2013Ap&SS.347..349L}. Figure~\ref{fig:gamma_flux_hist} shows the distribution of predicted $\gamma$-ray fluxes for our 67 sources. The majority of our sample falls just below \textit{Fermi}'s point-source sensitivity threshold (vertical dashed line) after 15 years of integration, with only a few sources in the shaded region that could be individually detectable. However, the intrinsic scatter in the x-ray/$\gamma$-ray correlation means we would expect some fraction of sources to scatter above the detection threshold if they followed a BL Lac–like distribution, and some fraction of sources from above the distribution to scatter below. Given the larger number of sources below the detection threshold, this dispersion would tend to lead towards a net increase in the number of detected sources. The absence of any firm detections therefore suggests that either the correlation breaks down in the low-mass regime, or that the SSC $\gamma$-ray emission in these systems---if it exists, is suppressed---potentially due to differences in jet orientation, accretion physics, source environments or enhanced internal/external $\gamma\gamma$ absorption. From the joint‑likelihood non‑detection, and adopting the BL Lac slope in Eq.~\ref{eq:x-ray-gamma-ray} ($\sim0.42)$, we constrain the normalization of the $F_{\gamma}-F_X$ relation to $\lesssim$-9.04 (95\% CL), indicating that dwarf AGN are at least a factor of $\sim$7 less efficient in producing $\gamma$-rays than BL Lacs at fixed x-ray luminosity.

\subsection{Joint-Likelihood Analysis}

We search for a cumulative $\gamma$-ray signal from the eROSITA dwarf-AGN sample with the joint likelihood analysis. For each source, we use per-bin likelihood scans $\Delta\log\mathcal{L}($d$N/$d$E; E_{\mathrm{ref}})$ above $E_{\rm min}=500$~MeV and evaluate a joint TS on a grid in photon index $\Gamma$ and a shared amplitude $A$. Source normalizations are tied as $c_i=Aw_i$, where we consider three physically motivated choices for the weights: (i) uniform ($w_i=1$), (ii) distance-only ($w_i \propto 1/d_i^2$), and (iii) mass-distance ($w_i \propto M^\alpha_{{\rm IMBH},i}/d^2$), where $\alpha \in \{1.0, 1.2, 1.5, 2\}$. Note that the case $\alpha=0$ reduces to (ii). To quantify significance, we compare it to 400 composite blank realizations per weighting scheme. Each blank realization is built by drawing 67 high-latitude blank fields and then assigning ``AGN-like'' physics weights via resampling of the $(M_{\rm IMBH}, d)$ pairs from the real sample.

\subsubsection{Potential Population-Level Soft Spectral Excess}
\label{subsubsec:soft_excess}

\begin{figure*}[t]
\centering
\includegraphics[width=0.32\textwidth]{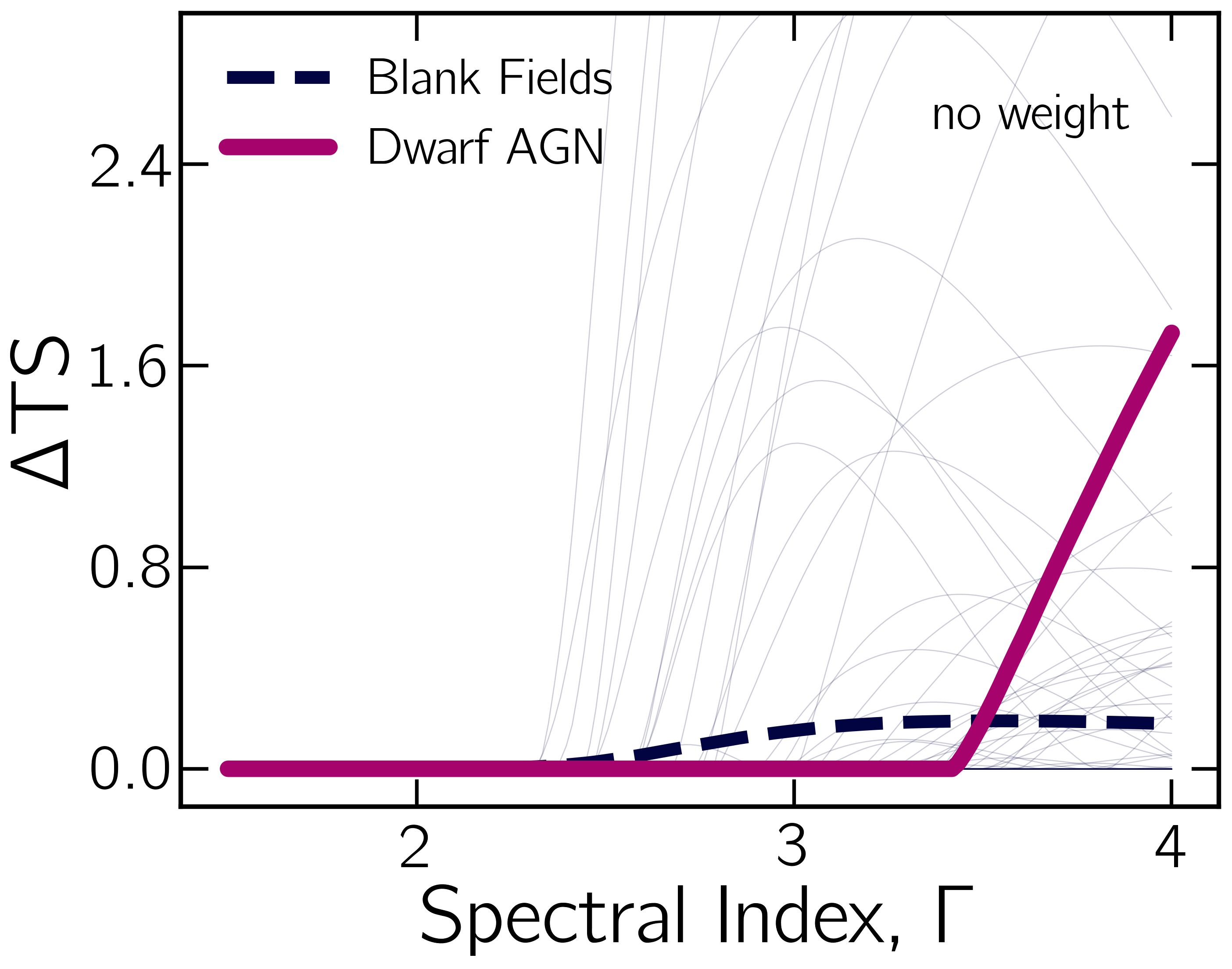}
\includegraphics[width=0.32\textwidth]{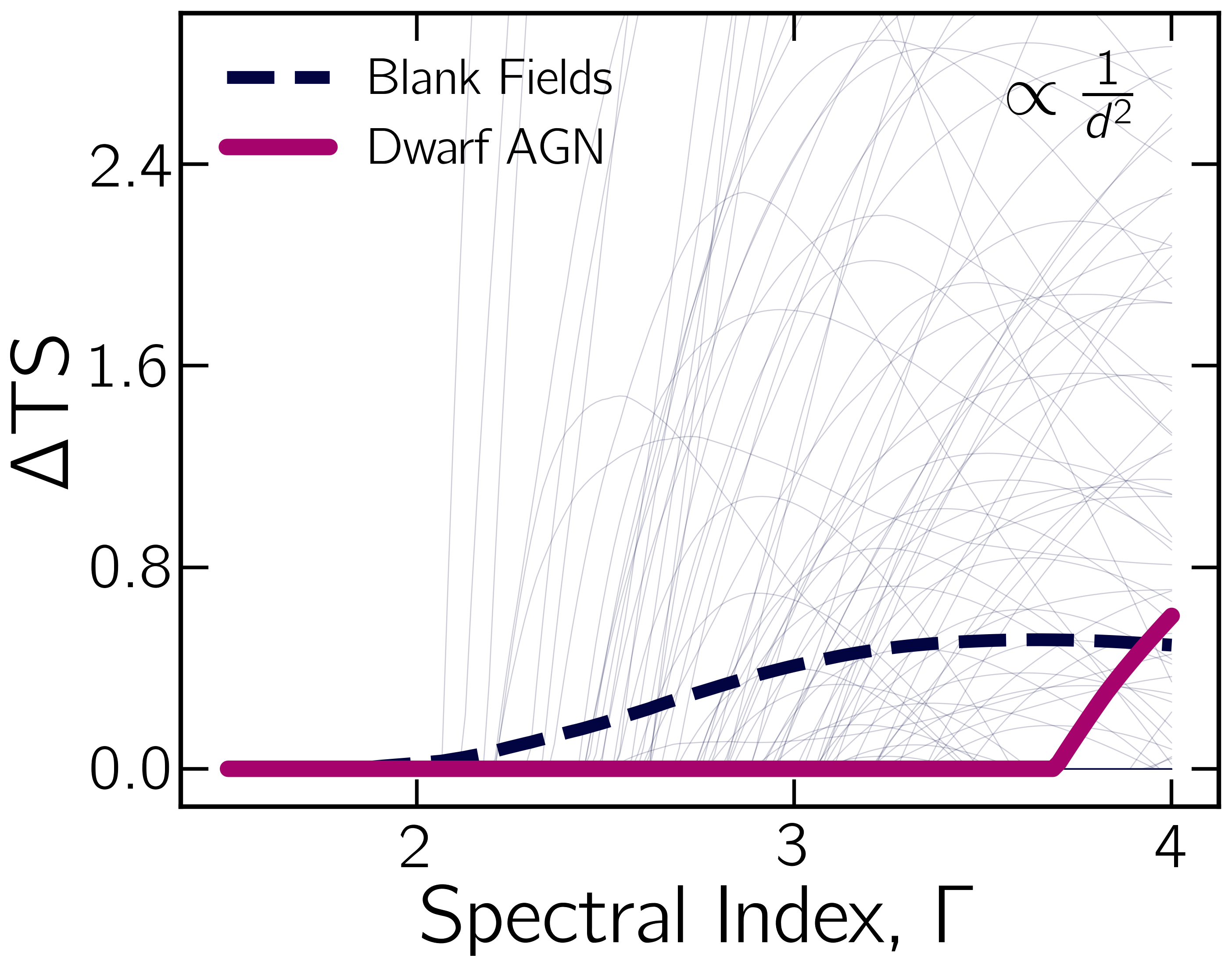}
\includegraphics[width=0.32\textwidth]{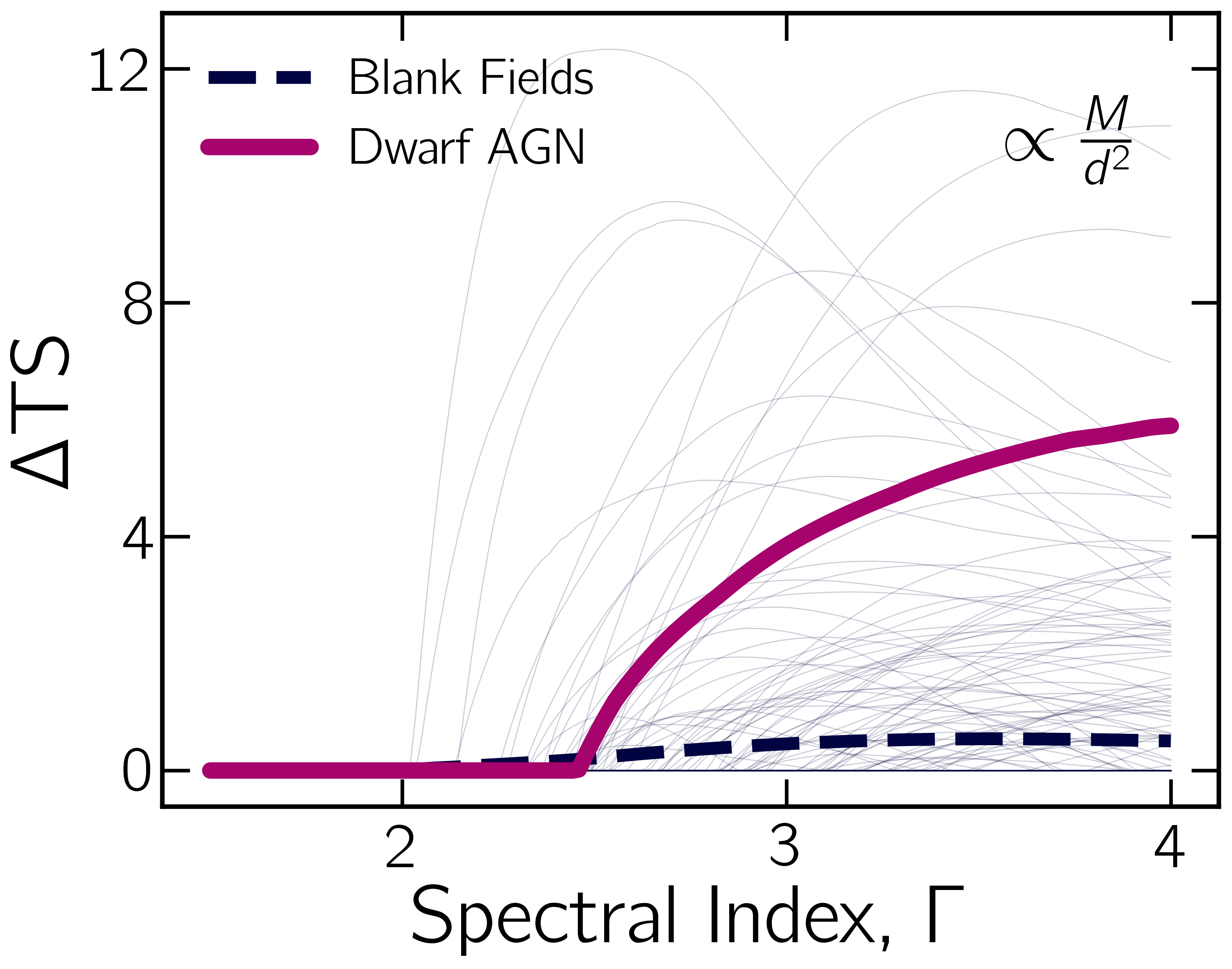}
\caption{TS profile as a function of spectral index $\Gamma$ for x-ray selected dwarf AGN across three weighting schemes. \textbf{Left:} Unweighted analysis giving equal contribution to all sources. \textbf{Center:} Distance-weighted analysis ($\propto 1/d^2$) accounting for geometric flux dilution. \textbf{Right:} Mass-distance weighted analysis ($\propto M_{\rm IMBH}/d^2$) testing dependence on IMBH mass. In all panels, the solid magenta line shows the dwarf AGN sample, the thin navy lines show individual joint likelihood analysis results from random samples of 67 blank fields, the dashed navy line shows the blank field mean. The modest soft spectral excess appears at $\Gamma >3$ in the mass-distance and uniform weighting schemes, while the distance-only weighting shows no significant excess above background.}
\label{fig:three_panel_TS}
\end{figure*}

Our joint likelihood analysis reveals a potential soft spectral excess at photon indices $\Gamma \gtrsim 3$ that appears most prominently in the mass-distance weighting scheme, with weaker evidence in uniform weighting, and no significant excess in distance-only weighting, as illustrated in Fig.~\ref{fig:three_panel_TS}. We summarize the significances of any potential detection by quantifying the (i) global (look-elsewhere-corrected) $p$-value, obtained by comparing the AGN's TS over the scanned $\Gamma$ grid to the distribution of TS values from the 400 blank realizations (i.e., trials-corrected significance for the $\Gamma$ scan); and (ii) the pointwise $p$-value evaluated at the AGN TS peak $\Gamma_{\rm peak}$. The results are summarized in Table~\ref{tab:soft_excess}.

\begin{table}[t!]
\centering
\caption{Significance of the soft spectral excess in the joint-likelihood analysis across weighting schemes. Columns 2--3 report empirical $p$–values from blank-field ensembles (pointwise at the peak photon index, $p_{\rm peak}$, and trials-corrected over the $\Gamma$ scan, $p_{\rm glob}$), with one-sided Gaussian equivalents in parentheses. Column 4 shows the Wilks/Poisson significance computed from the peak TS using the Chernoff boundary correction. We adopt the empirical values for interpretation.}
\label{tab:soft_excess}
\begin{tabular}{lccc}
\hline\hline
\textbf{Weighting Scheme} & $p_{\rm peak}$ ($\sigma$) & $p_{\rm glob}$ ($\sigma$) & Wilks ($\sigma$)\\
\hline
$M_{\rm IMBH}/d^2$ & 0.0125 (2.24) & 0.0250 (1.96) & 2.43 \\
Uniform                              & 0.0275 (1.90) & 0.0325 (1.85) & 1.32 \\
$1/d^{2}$           & 0.1325 (1.11) & 0.1550 (1.02) & 0.78 \\
\hline
\end{tabular}
\end{table}

The mass--distance weighting scheme ($M_{\rm IMBH}/d^2$) yields the highest significance, with a global signal of 2.0$\sigma$. The uniform weighting shows a weaker signal (1.9$\sigma$ global), while the distance-only weighting is consistent with background expectations (1.0$\sigma$ global). None of these signals reach the conventional 5$\sigma$ threshold. While the mass--distance and uniform weighting schemes both show a preference for very soft spectra at $\Gamma > 3$, the modest significances and the absence of a signal in distance-only weighting suggest these results should be interpreted cautiously.

\subsubsection{IMBH Mass Dependence of $\gamma$-Ray Emission}
Figure~\ref{fig:alpha_scan} summarizes the joint-likelihood scan in which source weights scale as $w_i \propto M_i^{\alpha}/d_i^{2}$ for representative $\alpha$ values. A coherent, very soft excess appears at $\Gamma \gtrsim 3$ whenever the weighting includes an approximately linear mass factor: the profiles for $\alpha=1$, $1.2$, and $1.5$ all rise smoothly and peak near $\Gamma \sim 3.8$--$4.0$, with similar shapes and amplitudes.  Using the blank ensembles to calibrate significance (as in Sec.~\ref{subsubsec:soft_excess}), the near-linear cases ($\alpha \simeq 1$--$1.5$) achieve preferences at the $\sim$2$\sigma$ level, comparable to the $M/d^2$ result quoted in Table~\ref{tab:soft_excess}, while $\alpha=0$ is consistent with the background. Very steep scaling with $\alpha= 2$ shows no improvement over blanks, suggesting that the signal is not dominated by only the most massive black holes in our sample. The peak position in $\Gamma$ is stable across all mass scalings—suggesting that the softness of the putative signal is not driven by the choice of $w_i$, whereas the amplitude is maximized for near-linear mass weights.

\section{Discussion \& Conclusions}
\label{sec:discussion}

Our $\gamma$-ray analysis of 74 x-ray selected dwarf AGN using 15 years of \textit{Fermi}-LAT data reveals no individual detections above the standard TS $>25$ threshold. The joint-likelihood analysis shows a modest but consistent preference for \emph{very soft} spectra, peaking at $\Gamma\simeq3.8$--$4.0$. The preference is most pronounced under $M_{\rm IMBH}/d^2$ weighting, where the local significance reaches $p_{\rm peak}=0.0125$ ($2.2\sigma$) and the global significance after accounting for the $\Gamma$ scan remains $p_{\rm glob}=0.025$ ($2.0\sigma$). Uniform weighting yields a slightly weaker excess (global $\sim1.9\sigma$), while distance-only weighting is fully consistent with background expectations ($\sim1.0\sigma$). These results suggest---if taken at face value---that any putative population component would scale with black hole mass (rather than, e.g., distance alone). 

The $\alpha$-scan clarifies that the putative population component is not driven solely by geometric weighting: the amplitude of the soft excess is largest for near-linear mass scalings ($\alpha\sim 1$--$1.5$) and diminishes for both distance-only (\(\alpha=0\)) and strongly super-linear (\(\alpha=2\)) weights (Fig.~\ref{fig:alpha_scan}). The peak location in $\Gamma$ is essentially unchanged across values of $\alpha$, indicating that the softness is a property of the hint signal rather than of the weighting prescription. Interpreted phenomenologically, this favors scenarios in which the $\gamma$-ray emission tracks black-hole mass roughly linearly at fixed Eddington ratio (Table~\ref{tab:alpha_summary}). However, the overall preference remains modest after accounting for the $\Gamma$ scan, and scanning $\alpha$ introduces an additional trials factor.

At this stage, we do not favor any single physical model over another; rather, we speculate on several potential explanations for the observed soft spectral behavior. We stress, however, that the global significances remain $\lesssim 2\sigma$ in all cases, so these are \emph{hypotheses}, not claims. %Read cautiously, the signal is strongest with a population component whose $\gamma$-ray amplitude increases with black-hole mass (rather than, e.g., distance alone). 

One possible interpretation is that the soft excess is driven by \textit{accretion-related processes}. In many AGNs, the innermost regions near the accretion disk and its hot corona can generate high-energy photons via mechanisms such as inverse Compton scattering. If such disk and corona emission dominates over jet-related activity in dwarf AGN, the resulting $\gamma$-ray spectrum would naturally appear softer \cite{Inoue:2019fil, Murase:2023ccp}. We note that typical eROSITA hardness ratios for these targets imply coronal x-ray photon indices $\Gamma_X \sim 1.7$--$2.0$ so a putative $\Gamma_{\gamma} \sim3.8$--$4.0$ component (above 500 MeV) would be consistent with non-beamed leptonic IC in cooler plasmas or with host cosmic-ray components, rather than with hard, beamed jet spectra.

Alternatively, the soft excess may arise if the AGN in dwarf galaxies are \textit{misaligned}. In this scenario, even if jets are present, they are not closely aligned with our line-of-sight. Without significant Doppler boosting, the emission from relativistic jets would be less prominent and the observed spectrum might be dominated by isotropic emission from the accretion flow, leading again to a softer observed spectrum. 

It is also possible that additional, non-AGN processes could contribute to the soft $\gamma$-ray emission. For instance, \textit{cosmic-ray interactions} in the host galaxy---such as those stemming from enhanced star formation or supernova activity---could yield a component of $\gamma$-rays via pion decay or inverse Compton scattering on ambient photon fields. Similarly, \textit{AGN-driven outflows or winds} interacting with the interstellar medium may create shocks that accelerate particles, imprinting a soft $\gamma$-ray spectrum.

Another intriguing---though more speculative---possibility is that the soft excess is partially influenced by \textit{dark matter-related processes}. While highly model-dependent, the annihilation or decay of a light dark matter component in regions of enhanced density near the central black hole might contribute $\gamma$-rays that are preferentially soft, thereby mimicking or enhancing the signature of accretion-powered emission \cite{1998SilkRees, Bertone:2005hw, Ferrer:2017xwm}.

At present, our analysis cannot distinguish among these scenarios. The lack of significant individual detections, combined with the tentative, population-level soft excess, points either to a subtle collective signal or to residual background/systematic fluctuations. Accordingly, we refrain from favoring a single interpretation. Both outcomes are scientifically informative: if real, the signal would imply that dwarf AGN differ from massive-galaxy AGN in their $\gamma$-ray spectra, potentially highlighting different emission channels in the IMBH regime; if not, the null result constrains the efficiency of $\gamma$-ray production in low-mass systems relative to their x-ray emission.

Future work can sharpen these conclusions along several directions. Observationally, deeper \textit{Fermi}-LAT exposures and eventual next-generation $\gamma$-ray telescopes at MeV energies will provide critical sensitivity to probe below the current limits. Multiwavelength follow-up---including x-ray spectral diagnostics of accretion states, radio constraints on jet power and orientation, and infrared tracers of star formation---will help separate AGN-related from host-galaxy contributions. A valuable extension of this work would involve constructing matched control samples of star-forming dwarf galaxies without AGN activity, which could directly isolate any host-galaxy contribution to the observed soft $\gamma$-ray hint and test whether the signal correlates with black hole mass or alternative galaxy properties such as stellar mass, star formation rate, or gas content. On the theoretical side, dedicated modeling of $\gamma$-ray emission in the IMBH regime, as well as predictions for dark matter signatures in dwarf AGN environments, will be essential to interpret any confirmed excess. 
\vspace{-0.07cm}

To conclude, while no firm detection is found in this paper, our results point to intriguing hints of a very soft $\gamma$-ray component in dwarf AGN. Confirming or refuting this possibility will require coordinated observational and theoretical efforts, with implications for both high-energy astrophysics and the physics of IMBHs.

\acknowledgements

We thank Vivienne Baldassare and Akaxia Cruz for helpful discussions, and Andrea Sacchi for providing the list of excluded sources. We also acknowledge the independent work of Rodrigo Nemmen and his group on this topic, which has not yet been published, and appreciate their openness in sharing insights during TeVPA 2024 in Chicago. MC and TL acknowledge support from the Swedish Research Council under contract 2022-04283 and the Swedish National Space Agency under contract 117/19. TL also acknowledges sabbatical support from the Wenner-Gren foundation under contract SSh2024-0037. AHGP acknowledges support from the National Science Foundation under Grant No. AST-2008110.

This project made use of \texttt{Astropy} \cite{astropy:2022}, \texttt{Numpy} \citep{numpy:2020}, \texttt{Matplotlib} \cite{matplotlib:2007}, and \texttt{Pandas} \cite{pandas:2020} Python packages. We acknowledge the use of public data from the \textit{Fermi Science Support Center} data archive. We additionally acknowledge the use of OpenAI's ChatGPT to assist with editing and proofreading of the manuscript.

\bibliographystyle{apsrev4-1}

\appendix

\section{Excluded Sources from \citet{Sacchi:2024yic}.}
\label{app:excluded}

We also analyze sources excluded by \citet{Sacchi:2024yic}, including ULXs, background AGN, and XRBs. This analysis serves as a control sample to validate that our dwarf AGN population exhibits distinct $\gamma$-ray properties compared to known contaminants. These objects are passed through the same $\gamma$-ray pipeline as the main AGN sample to determine their TS distribution and probe population-specific effects. This helps quantify how sources with similar x-ray properties might bias the significance of any detected signal in our AGN sample, addressing potential selection effects. Excluded sources are listed in Table~\ref{tab:excluded_sources} and their sky distribution is shown in Fig.~\ref{fig:blanks}.

Following the procedure of the original dwarf AGN analysis, we exclude all sources within $|b| < 15^\circ$ to mitigate contamination from diffuse Galactic emission (PGC017603, ESO125$-$001, ESO495$-$021, IC2596, PGC090183). We additionally remove sources located too close to bright $\gamma$-ray emitters. In particular, we exclude PGC2828331, which lies only $0.01^\circ$ from the known FSRQ PKS1510$-$089 (4FGLJ1512.8$-$0906). PKS1510$-$089, at $z=0.361$, is one of the brightest and most variable $\gamma$-ray blazars on the sky. It has been repeatedly detected at both GeV and TeV energies with instruments including \textit{Fermi}-LAT, H.E.S.S., and MAGIC, exhibiting multiwavelength flares and even sub-hour variability at very-high energies. We thus remove it from our analysis. 
\begin{table}[H]
\centering
\caption{Excluded sources analyzed for comparison, indicating their nature (e.g., ULX: Ultraluminous X-ray source, XRB: X-ray binary, AGN: Active Galactic Nucleus). 
Sources not used in our analysis are shown in \textit{italics}. The final excluded-source control sample consists of 40 objects.}
\label{tab:excluded_sources}
\begin{tabular}{lccc}
\hline
\hline
\textbf{Object Name} & \textbf{RA} & \textbf{Dec} & \textbf{Flag} \\
\hline
6dFJ0243580-091424 & 40.99 & -9.24 & ULX \\
PGC715598          & 46.33 & -30.58 & ULX \\
PGC011744 & 47.08 & -13.90 & AGN \\
NGC1345            & 52.38 & -17.78& XRB \\
PGC013036          & 52.67 & -54.58 & XRB \\
PGC013821          & 56.66 & -16.55 & XRB \\
PGC014118          & 58.94 & -42.37 & ULX \\
NGC1518            & 61.71 & -21.17 & AGN \\
ESO118-019         & 64.75 & -58.26 & XRB \\
PGC015573          & 68.76 & -14.23 & XRB \\
PGC016320          & 73.75 & -37.26 & AGN \\
NGC1796            & 75.68 & -61.14 & XRB \\
\textit{PGC017603}          & 85.43 & 6.68  & ULX \\
PGC828863          & 89.55 & -21.39 & ULX \\
PGC571171          & 103.50 & -41.86 & ULX \\
ESO162-017         & 108.98 & -57.34 & XRB \\
\textit{ESO125-001}         & 128.81 & -58.61 & ULX \\
\textit{ESO495-021 }        & 129.06 & -26.41 & XRB \\
PGC024880          & 132.86 & 39.59  & AGN \\
PGC3090323         & 133.63 & 17.63  & AGN \\
UGC05588           & 155.24 & 25.36  & XRB \\
\textit{IC2596}             & 158.55 & -73.24 & XRB \\
ESO501-092         & 161.41 & -23.01 & AGN \\
NGC3513            & 165.94 & -23.25 & XRB \\
PGC036612          & 176.36 & -10.06 & AGN \\
NGC4038            & 180.47 & -18.87 & ULX \\
PGC038055          & 180.74 & -20.93 & AGN \\
GAMA099049         & 180.91 & 1.04   & ULX \\
NGC4116            & 181.90 & 2.69   & XRB \\
2MASXJ12154983+2749367 & 183.96 & 27.83 & ULX \\
SDSSJ121855.02+142445.5 & 184.73 & 14.41 & ULX \\
LAMOSTJ122142.12+250610.5 & 185.43 & 25.10 & AGN \\
NGC4670            & 191.32 & 27.13  & AGN \\
PGC3105909         & 193.27 & -3.22  & AGN \\
NGC4765            & 193.31 & 4.46   & ULX \\
PGC045512          & 196.99 & -16.69 & ULX \\
ESO270-017         & 203.70 & -45.55 & XRB \\
NGC5253            & 204.98 & -31.64 & AGN \\
\textit{PGC090183}          & 207.08 & -50.98 & XRB \\
SDSSJ140059.02+120637.5 & 210.25 & 12.11 & AGN \\
NGC5595            & 216.06 & -16.72 & XRB \\
PGC3113751         & 226.73 & 0.18   & AGN \\
\textit{PGC2828331}         & 228.21 & -9.10  & AGN \\
ESO338-004         & 291.99 & -41.58 & ULX \\
PGC2793893         & 339.55 & -39.67 & AGN \\
6dFJ2335233-394505 & 353.85 & -39.75 & AGN \\
\hline
\hline
\end{tabular}
\end{table}

\begin{figure}[H]
\centering
\includegraphics[width=0.45\textwidth]{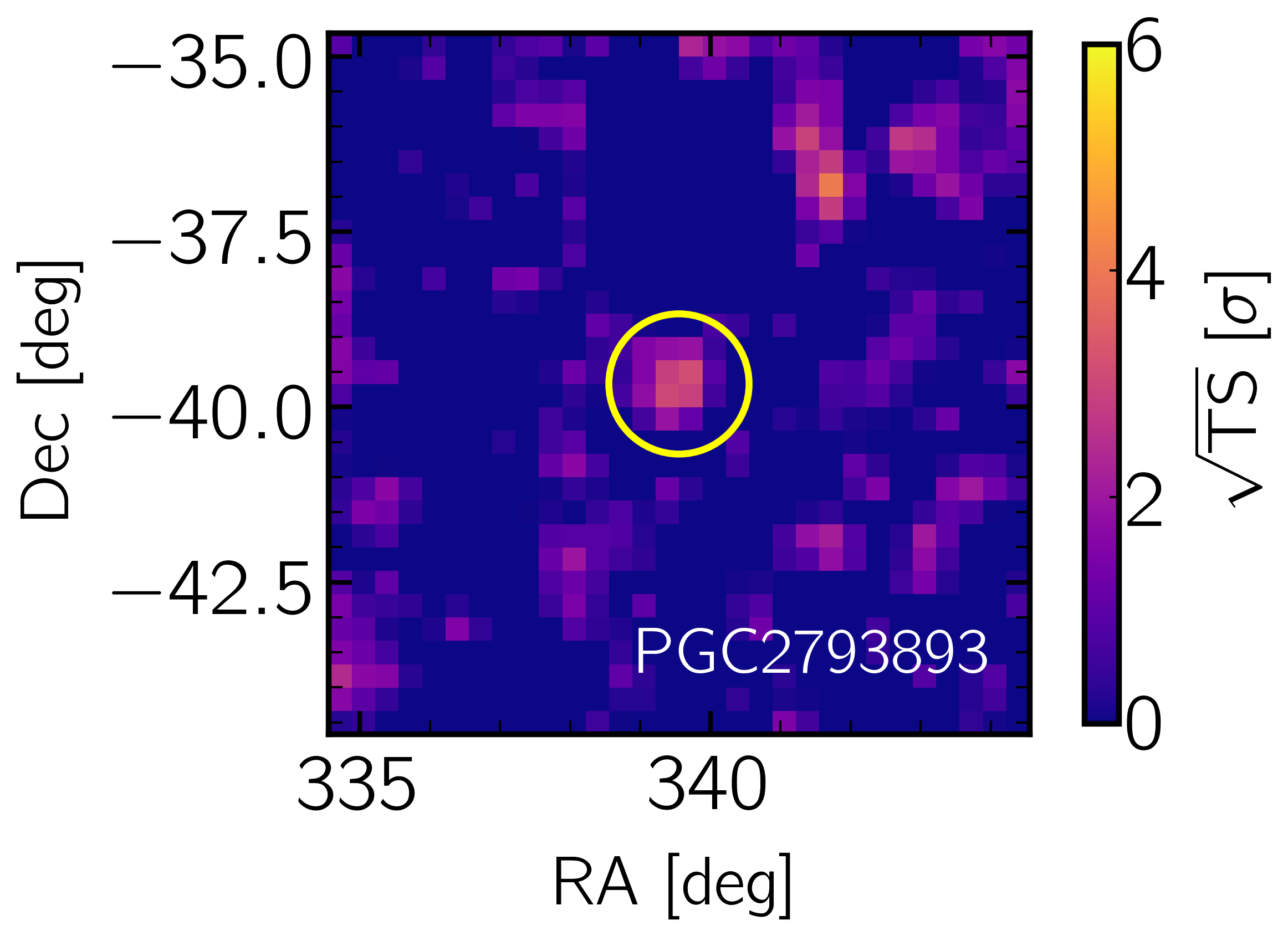}
\caption{Residual $\sqrt{TS}$ map for the PGC2793893 field. The significant excess ($\sqrt{TS} \sim 4$) at the source position (yellow circle) confirms $\gamma$-ray emission coincident with the BL Lac object 2MASX~J22381274-3940195.}
\label{fig:pgc2793893_tsmap}
\end{figure}

\textbf{Note on PGC2793893.} --- Albeit not listed  in 4FGL-DR4 \cite{Fermi-LAT:2022byn}, PGC2793893 shows significant $\gamma$-ray emission ($\sim 4\sigma$) with a hard spectral index ($\Gamma \sim 2$), characteristic of blazar-type emission. Cross-referencing with the SIMBAD astronomical database\footnote{\url{https://simbad.cds.unistra.fr/simbad/}, accessed on September 2, 2025.} reveals that this position coincides with 2MASX~J22381274-3940195, a confirmed BL Lac object at redshift $z = 0.250$ classified in multiple blazar catalogs \citep{2015Ap&SS.357...75M, Massaro:2008ye, Chang:2019vfd}. Figure~\ref{fig:pgc2793893_tsmap} shows the residual $\sqrt{TS}$ map when this source is left unmodeled, confirming significant $\gamma$-ray excess at the source position. Given its relatively strong individual detection and potential to dominate any population-level signal, we exclude PGC2793893 from the joint likelihood analysis. This weak detection, however, validates exclusion criteria in Ref.~\cite{Sacchi:2024yic}, as the hard $\gamma$-ray spectrum ($\Gamma \sim 2$) contrasts sharply with the soft spectral excess hint ($\Gamma \sim 4$) observed in our dwarf AGN population, potentially alluding to fundamentally different $\gamma$-ray emission mechanisms compared between BL Lacs and dwarf AGN.

To further validate the robustness of our population-level results, we perform a bootstrap resampling of the excluded source sample. Specifically, we construct 100 bootstrap realizations by randomly drawing---with replacement---subsets of 67 sources (corresponding to the final sample after Galactic-plane and bright-neighbor cuts), providing an estimate of the variance in our joint-likelihood TS distribution and ensuring that any observed spectral property is not dominated by a small subset of outliers but is instead a population-level feature. 

\begin{figure*}
\centering
\includegraphics[width=0.32\textwidth]{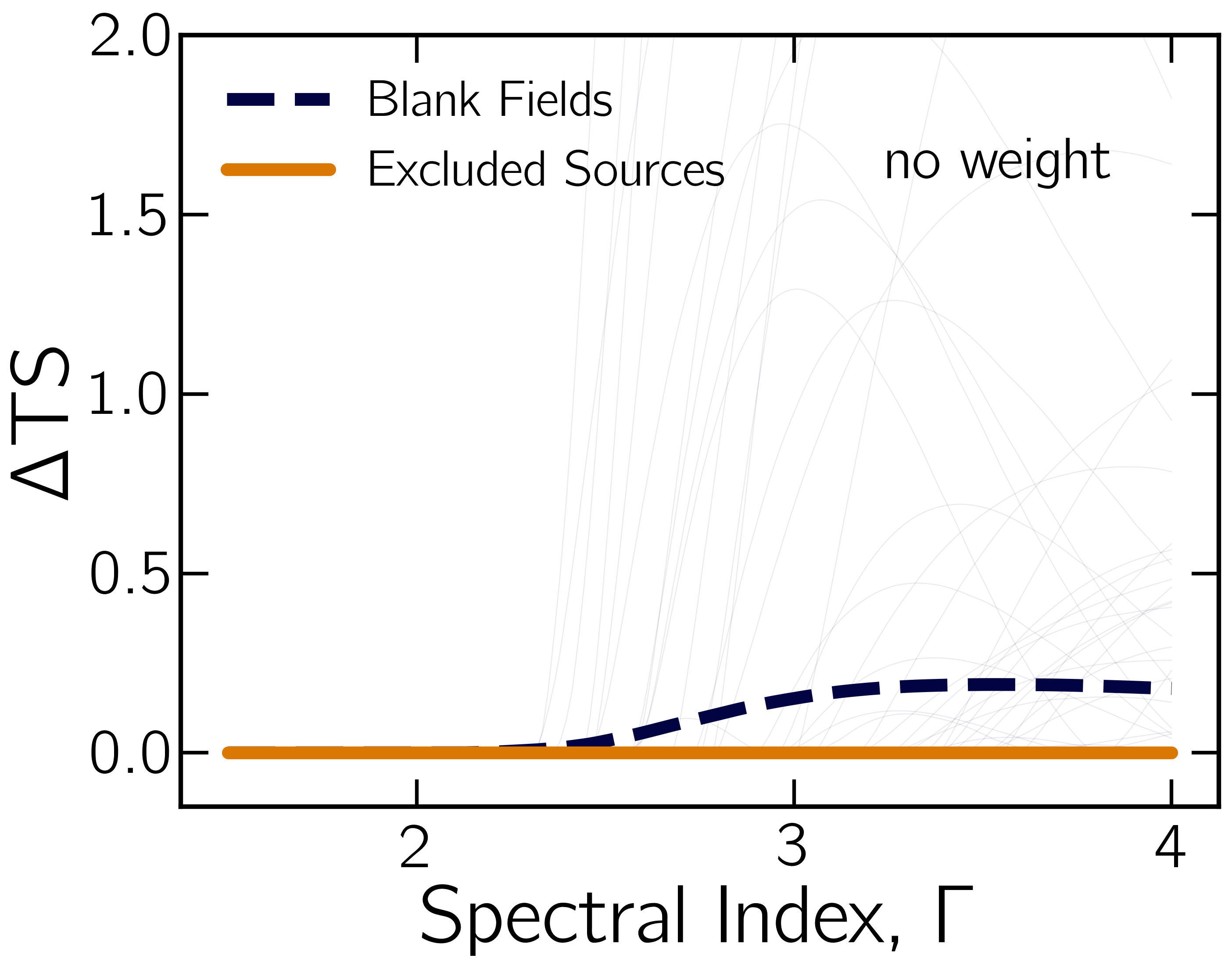}
\includegraphics[width=0.32\textwidth]{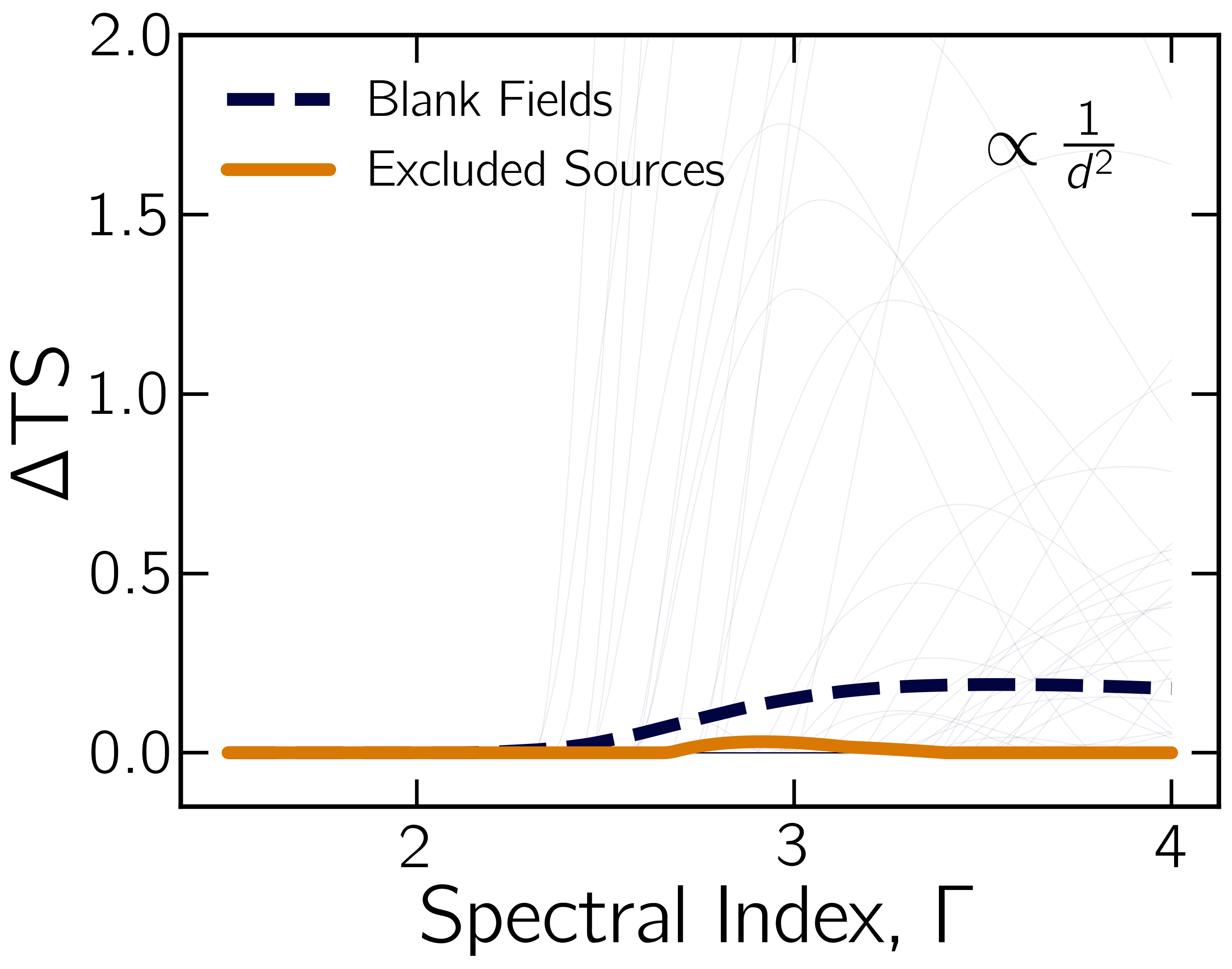}
\includegraphics[width=0.32\textwidth]{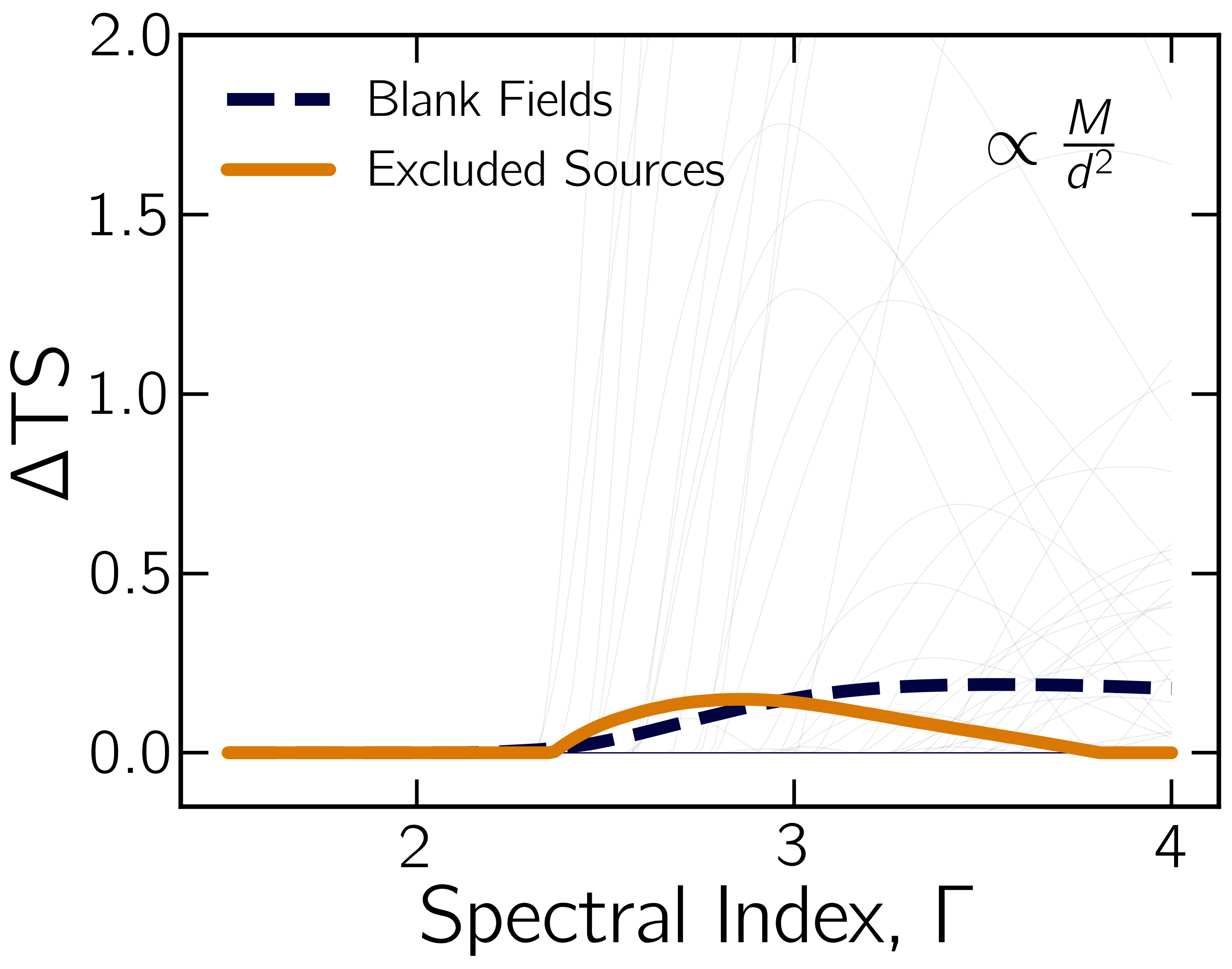}
\caption{The bootstrapped joint-likelihood TS profiles of excluded sources (orange) compared with the mean of the blank-field realizations (dashed navy) for different weighting schemes: (\textit{left}) uniform weighting, (\textit{center}) distance-only weighting ($1/d^{2}$), and (\textit{right}) mass–distance weighting ($M/d^{2}$). Excluded sources are consistent with the blank fields.}
\label{fig:excluded_vs_blank}
\end{figure*}

Figure~\ref{fig:excluded_vs_blank} compares the joint-likelihood TS profiles of the excluded-source control sample with those of the blank-field realizations across all weighting schemes. In all cases, the excluded sources are statistically consistent with the blank-field expectation and show no evidence for a soft excess. This confirms that the $\gamma$-ray properties of ULXs, XRBs, and background AGN excluded from the dwarf AGN sample do not mimic the tentative soft-spectrum signal observed in our main analysis. Finally, the bootstrap scatter is at the $\Delta$TS~$\lesssim 0.5$ level, indicating minimal variance and that no subset of excluded sources drive the result.

\section{Analysis down to 100 MeV}
\label{app:100MeV}

To test the robustness of our soft spectral excess in Fig.~\ref{fig:alpha_scan}, we extend our analysis down to 100~MeV despite the increased systematic uncertainties at these energies. If the excess were purely due to instrumental effects at the 500~MeV threshold, we might expect it to disappear or change character when including the more systematics-dominated low-energy data. This is particularly true to the soft-spectrum of our best-fit model, which indicates that there should be a significantly higher number of photons in the 100--500~MeV range than throughout the rest of our analysis window. However, if astrophysical in origin, the spectral shape should remain consistent (though with inflated significance due to systematic effects).

Analyses below 500~MeV are dominated by well-documented systematic uncertainties in the \textit{Fermi}-LAT data \cite{Fermi-LAT:2013sme, Hooper_2016, Principe:2018lzn, Fermi-LAT:2016zaq}. The instrument response functions carry systematic uncertainties of 10-20\% at 100 MeV compared to ~5\% at energies near 1~GeV \cite{2009PhRvL.103y1101A}. Moreover, the angular resolution degrades dramatically from sub-degree precision above 1~GeV to $\sim3^\circ$ at 100~MeV, causing source confusion and spurious extended source detections. Energy reconstruction becomes less reliable below 100 MeV, with the energy resolution degrading from ~10\% at GeV energies to $\sim$20--30\% at 100~MeV \cite{2009ApJ...697.1071A}.
From a modeling perspective, astrophysical backgrounds, particularly of Galactic diffuse emission, become fundamentally unreliable at low energies with systematic uncertainties of 15-30\% \cite{Fermi-LAT:2014ryh}. These systematic effects, thus, can easily exceed statistical uncertainties and create apparent detections that are purely instrumental artifacts.

\begin{figure*}
\centering
\includegraphics[width=0.32\textwidth]{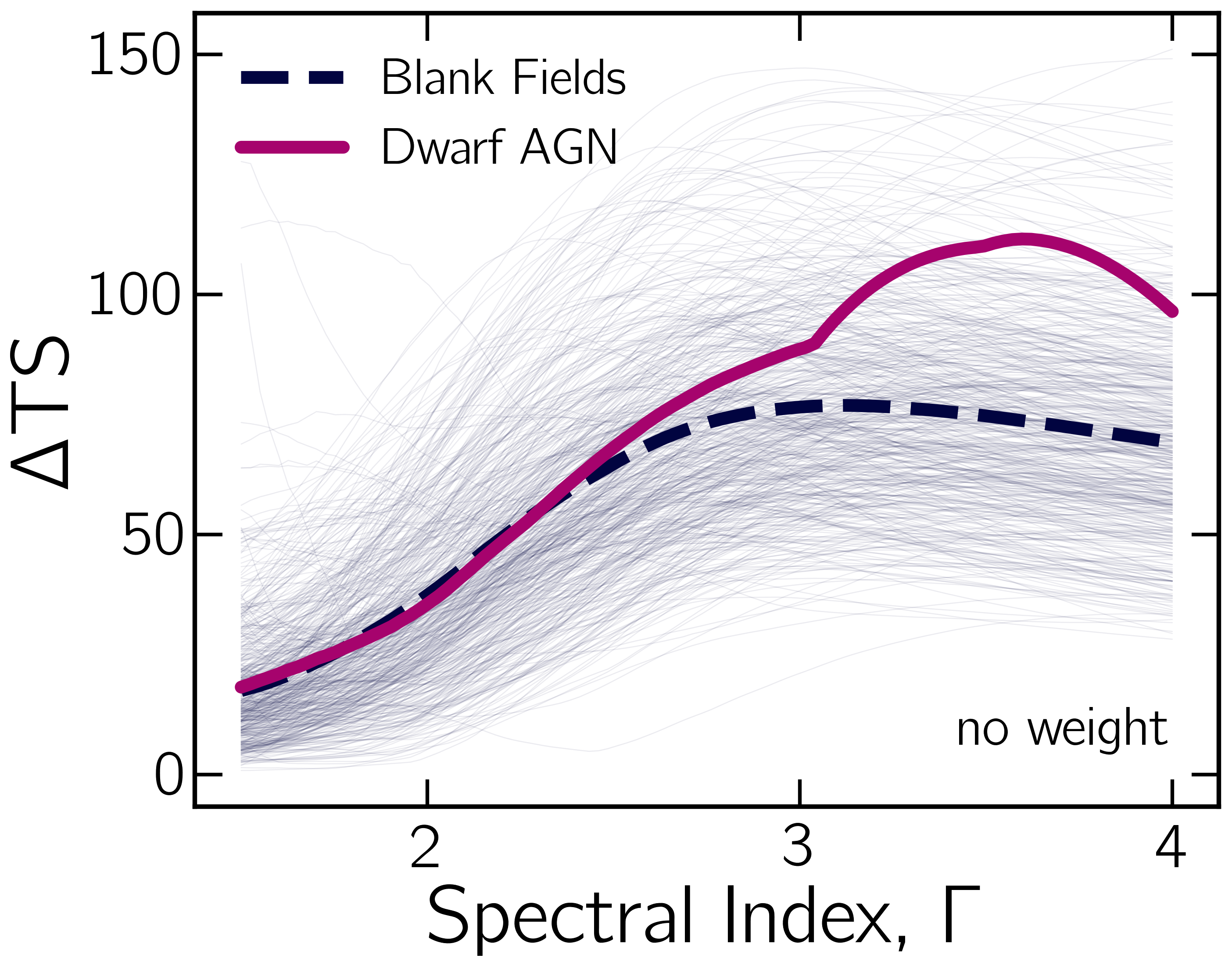}
\includegraphics[width=0.32\textwidth]{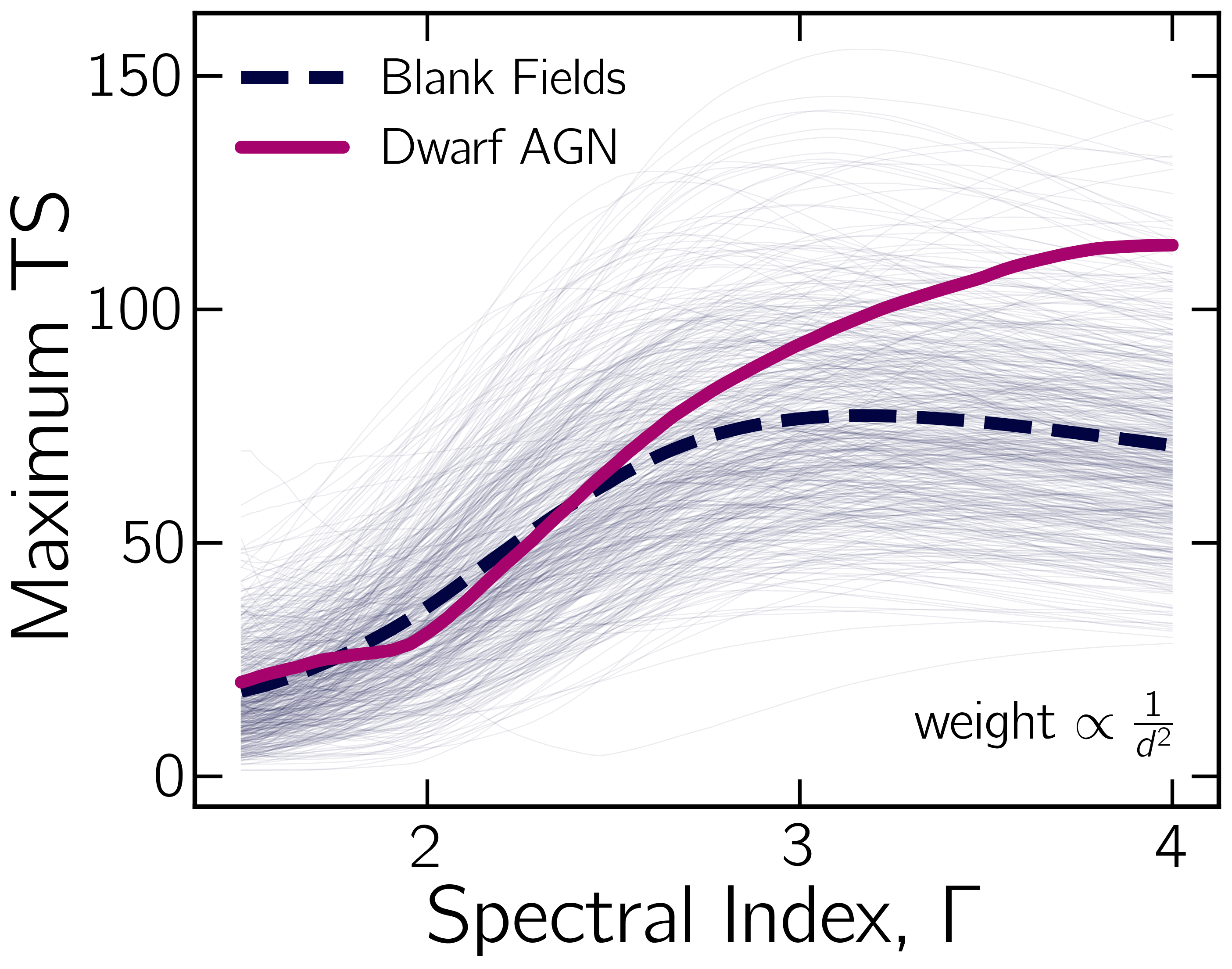}
\includegraphics[width=0.32\textwidth]{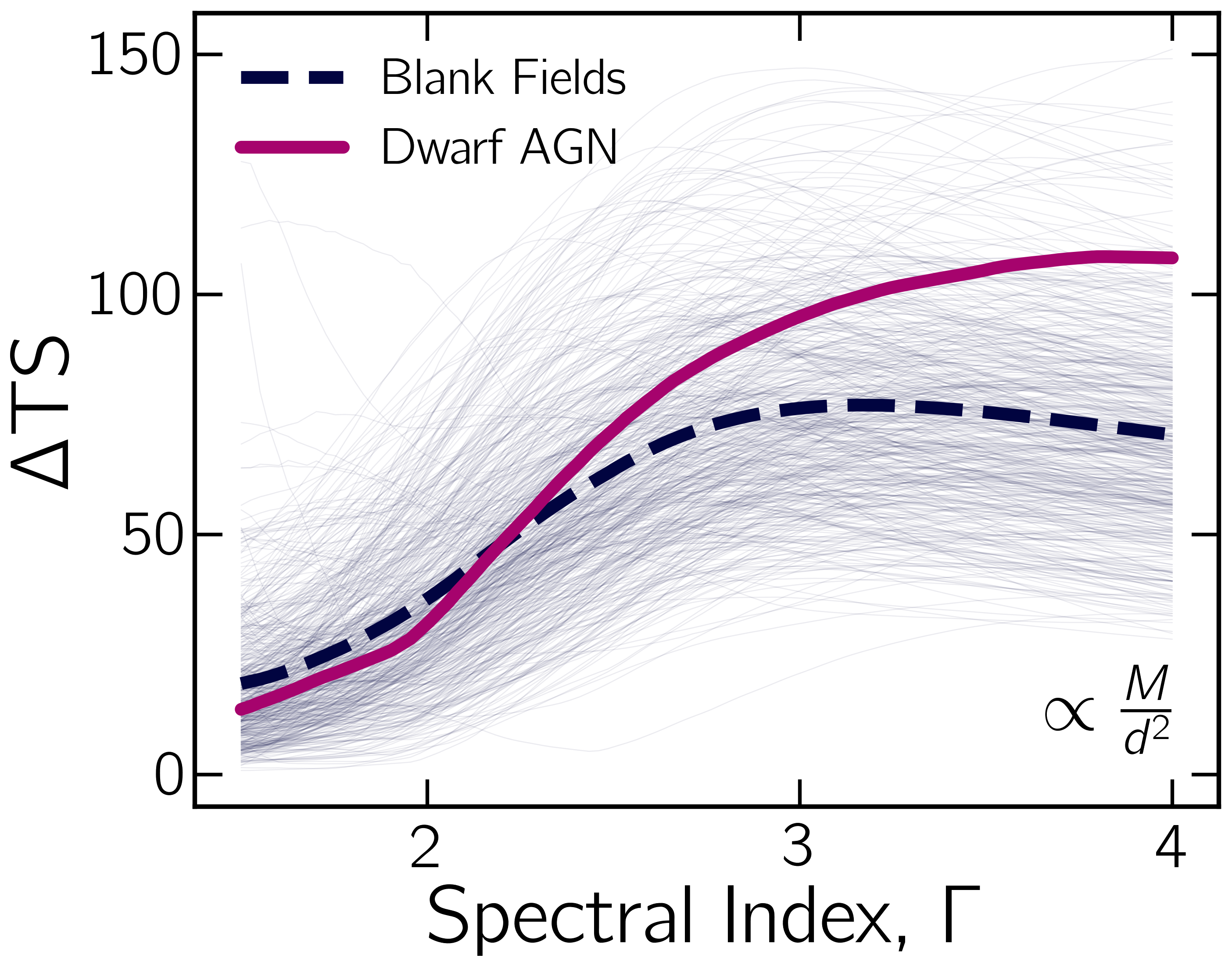}
\caption{Joint TS profile down to 100~MeV as a function of spectral index $\Gamma$ for x-ray selected dwarf AGN for different weighting schemes: (\textit{left}) uniform weighting, (\textit{center}) distance-only weighting ($1/d^2$), and (\textit{right}) mass-distance weighting ($M_{\rm IMBH}/d^2$). The solid magenta line shows the dwarf AGN sample, dashed navy line shows blank field mean. A modest soft excess at $\Gamma \sim 3.9$--$4.0$ is most prominent in the mass-distance weighted case.}
\label{fig:three_panel_TS_100MeV}
\end{figure*}

When we extend the joint likelihood analysis to down to 100~MeV, we retain the enhanced significance in the soft spectral regions, with spectral indices remaining around $\Gamma \simeq 3.8-4.0$ but with significantly inflated test statistics, shown in Fig.~\ref{fig:three_panel_TS_100MeV}. The analysis yields approximately $\sim2.4\sigma$ significance compared to blank fields, representing a modest but notable consistency with our 500~MeV analysis. We note that, for the mass-distance weighting scheme, the best-fit flux above 500~MeV deduced using the best-fit parameters from the 100-MeV analysis agrees with the direct 500~MeV fit at the $\sim$10\% level, well within the expected joint-likelihood uncertainties, further supporting consistency across thresholds. The consistency of the spectral shape across energy thresholds may suggest the signal may have some astrophysical origins rather than being purely systematic, though the enhanced significance likely stems from instrumental effects rather than of any real signal.

This comparison indicates that while our primary 500~MeV analysis shows only modest evidence for soft emission, it appears more robust than typical systematic artifacts. However, the fundamental systematic limitations of pair-conversion telescopes in the MeV range underscore the critical need for next-generation instruments designed specifically for MeV astronomy---such as AMEGO \cite{AMEGO:2019gny}, AMEGO-X \cite{Caputo:2022xpx}, e-ASTROGAM \cite{e-ASTROGAM:2017pxr}, or VLAST \cite{Pan:2024adp}---to definitively resolve the astrophysical nature of $\gamma$-ray emission from IMBH systems.

\end{document}